%% file: main.tex
\documentclass{LMCS}

\def\doi{8 (2:18) 2012}
\lmcsheading%
{\doi}
{1--42}
{}
{}
{Oct.~19, 2011}
{Jun.~28, 2012}
{}

\usepackage{enumerate}
\usepackage{hyperref}

\usepackage{graphicx}
\usepackage{wrapfig}
\usepackage{algorithmic}
\usepackage{algorithm}
\usepackage{pstricks}
\usepackage{pst-tree,pst-node}
\usepackage{proof}

\theoremstyle{plain}
\newtheorem*{ThmStar}{Theorem}
\newtheorem*{LemStar}{Lemma}

\input{macros}

\newpsobject{showgrid}{psgrid}{subgriddiv=1,griddots=10,gridlabels=6pt}

\begin{document}

\title[Explicit-Scheduler Symbolic-Thread]{
Software Model Checking with Explicit Scheduler and Symbolic Threads
}

\author[A.~Cimatti]{Alessandro Cimatti}
\author[I.~Narasamdya]{Iman Narasamdya}
\author[M.~Roveri]{Marco Roveri}
\address{Fondazione Bruno Kessler} 
\email{\{cimatti,narasamdya,roveri\}@fbk.eu}  





\begin{abstract}

In many practical application domains, the software is organized into
a set of threads, whose activation is exclusive and controlled by 
a cooperative scheduling policy: threads execute, without any interruption,
until they either terminate or yield the control explicitly to
the scheduler.


The formal verification of such software poses significant challenges.
On the one side, each thread may have infinite state space, and might
call for abstraction.  On the other side, the scheduling policy is
often important for correctness, and an approach based on abstracting
the scheduler may result in loss of precision and false
positives. Unfortunately, the translation of the problem into a purely
sequential software model checking problem turns out to be highly
inefficient for the available technologies.

We propose a software model checking technique that exploits the
intrinsic structure of these programs.
Each thread is translated into a separate sequential program and
explored symbolically with lazy abstraction, while the overall
verification is orchestrated by the direct execution of the scheduler.
The approach is optimized by filtering the exploration of
the scheduler with the integration of partial-order reduction. 

The technique, called ESST (Explicit Scheduler, Symbolic Threads) has
been implemented and experimentally evaluated on a significant set of
benchmarks. The results demonstrate that ESST technique is way
more effective than software model checking applied to the
sequentialized programs, and that partial-order reduction can lead to
further performance improvements.
\end{abstract}

\maketitle

\input{introduction}

\input{background}

\input{program-model}

\input{esst}

\input{po-for-esst}

\input{experiments}

\input{related-work}

\input{conclusions}

\bibliographystyle{alpha}
\bibliography{refs}

\newpage
\appendix
\input{appendix}

\end{document}

%% file: macros.tex
\usepackage{listings}
\usepackage{xspace}

\lstnewenvironment{codesnippet}[2]%
{%
\lstset{language=C++,
    morekeywords={sc_module,sc_event,sc_port,sensitive,%
    sc_in,SC_METHOD,SC_MODULE,SC_CTOR,SC_PROCESS,SC_METHOD,SC_THREAD,%
    SC_HAS_PROCESS,SC_ZERO_TIME,wait,notify,sc_start,sc_signal,%
    put,sc_export,sc_out,sc_interface,assert,main},
    backgroundcolor=\color{gray!10},
    keywordstyle=\color{blue!55}\bfseries,
    basicstyle=\tiny\sffamily,
    commentstyle=\color{green!60!black},
    stringstyle=\rmfamily,
    numbers=left,
    numberstyle=\tiny\ttfamily,
    numbersep=0.8em,
    showstringspaces=false,
    breaklines=false,
    frame=lines,
    float=!t,
    captionpos=b,
    framexleftmargin=2.5em,
    xleftmargin=2.5em,
    caption={#1},
    label={#2}
}%
}%
{}

\newcommand{\AND}{\wedge}

\newcommand{\IMPL}{\Rightarrow}
\newcommand{\IFF}{\Leftrightarrow}

\newcommand{\Int}{\mathbb{Z}}    
\newcommand{\setof}[1]{\{#1\}}               
\newcommand{\powerset}[1]{\mathcal{P}(#1)}   
\newcommand{\sem}[1]{\ensuremath{[\![#1]\!]}}   
\newcommand{\proj}[1]{\ensuremath{proj_{#1}}}   

\newcommand{\mainthread}{main}
\newcommand{\TCFG}[1]{G_{#1}}
\newcommand{\TLVar}[1]{\mathit{LVar}_{#1}}
\newcommand{\Var}{\mathit{Var}}
\newcommand{\Val}{\Int}
\newcommand{\GVar}{\mathit{GVar}}
\newcommand{\SVar}{\mathit{SVar}}
\newcommand{\State}{\mathit{State}}
\newcommand{\st}{s}
\newcommand{\ev}{ev}
\newcommand{\tst}[1]{\st_{#1}}
\newcommand{\tev}[1]{\ev_{#1}}
\newcommand{\gst}{gs}
\newcommand{\sst}{\mathbb{S}}
\newcommand{\StDom}[1]{\mathit{Dom}(#1)}
\newcommand{\cg}{\gamma}
\newcommand{\tcg}[1]{\cg_{#1}}
\newcommand{\conf}[3]{\langle #1, #2, #3 \rangle}
\newcommand{\ReachC}[1]{\mathit{Reach}(#1)}
\newcommand{\tstatus}[1]{st_{#1}}
\newcommand{\RUNNING}{\mathit{Running}}
\newcommand{\RUNNABLE}{\mathit{Runnable}}

\newcommand{\WAITING}{\mathit{Waiting}}

\newcommand{\SchedFn}{\mathit{Sched}}

\newcommand{\schedstates}{\mathit{SState}}
\newcommand{\nrschedstates}{\mathit{SState}_{No}}
\newcommand{\orschedstates}{\mathit{SState}_{One}}
\newcommand{\schedstset}{\mathcal{S}}
\newcommand{\primcalls}{\mathit{PrimitiveCall}}

\newcommand{\semE}[1]{\sem{#1}_{\mathcal{E}}}
\newcommand{\semB}[1]{\sem{#1}_{\mathcal{B}}}

\newcommand{\reg}{r}
\newcommand{\Preds}{\Pi}
\newcommand{\lprec}[1]{\pi_{#1}}
\newcommand{\tprec}[1]{\pi_{#1}}
\newcommand{\gprec}{\pi}
\newcommand{\fvar}[1]{\mathit{fvar}(#1)}
\newcommand{\Theory}{\mathcal{T}}
\newcommand{\spfun}{\mathit{SP}} 
\newcommand{\SPO}[1]{\spfun_{#1}}
\newcommand{\SP}[2]{\SPO{#2}(#1)}
\newcommand{\ASPO}[2]{\spfun_{#1}^{#2}}
\newcommand{\ASP}[3]{\ASPO{#2}{#3}(#1)}
\newcommand{\trans}{\rightarrow} 
\newcommand{\transsem}[1]{\stackrel{#1}{\trans}}

\newcommand{\transX}[1]{\stackrel{#1}{\trans}}
\newcommand{\transa}{\alpha} 
\newcommand{\transb}{\beta} 
\newcommand{\transc}{\gamma}
\newcommand{\transaS}[1]{\transa(#1)} 
\newcommand{\transaSS}[2]{\transa(#1,#2)} 
\newcommand{\StransaS}[2]{#1\transX{\transa}#2} 
\newcommand{\transbS}[1]{\transb(#1)} 
\newcommand{\transbSS}[2]{\transb(#1,#2)} 
\newcommand{\StransbS}[2]{#1\transX{\transb}#2}

\newcommand{\Terr}{\mathit{T_{err}}}
\newcommand{\Err}[2]{\mathit{E}_{#1,#2}}
\newcommand{\ABlock}{\mathit{ABlock}}

\newcommand{\ESST}{\textsc{ESST}\xspace}
\newcommand{\CEGAR}{\textsc{CEGAR}\xspace}
\newcommand{\POR}{\textsc{POR}\xspace}

\newcommand{\CFG}{CFG\xspace}
\newcommand{\CFGs}{CFGs\xspace}
\newcommand{\ART}{ART\xspace}
\newcommand{\ARTs}{ARTs\xspace}
\newcommand{\ARF}{ARF\xspace}

\newcommand{\havoc}{\ensuremath{\textsc{havoc}}\xspace}
\newcommand{\Sexec}{\ensuremath{\textsc{Sexec}}\xspace}
\newcommand{\Sched}{\ensuremath{\textsc{Sched}}\xspace}
\newcommand{\Persistent}{\textsc{Persistent}\xspace}
\newcommand{\Sleep}{\textsc{Sleep}\xspace}
\newcommand{\code}[1]{\texttt{\small #1}}

\newcommand{\ststate}[2]{\langle #1,#2 \rangle}
\newcommand{\arfpath}[2]{\mathit{ARFPath}(#1,#2)} 
\newcommand{\nonrunning}[1]{\mathit{NonRunning}(#1)} 
\newcommand{\tedge}{\varepsilon}
\newcommand{\tpath}{\rho}
\newcommand{\fpath}{\hat{\rho}}
\newcommand{\fconn}{\kappa}
\newcommand{\opsseq}{\sigma}
\newcommand{\pathops}[1]{\opsseq_{#1}}
\newcommand{\suppath}[1]{\mathit{sup}(#1)}
\newcommand{\frnode}{\eta}
\newcommand{\forest}{\mathcal{F}}
\newcommand{\NodesF}[1]{\mathit{Nodes}(#1)}
\newcommand{\simtrans}[1]{\stackrel{#1}{\trans}}

\newcommand{\enabled}[1]{\mathit{enabled}(#1)}

\newcommand{\Reach}[2]{\mathit{Reach}(#1,#2)}
\newcommand{\ReachRed}[2]{\mathit{Reach}_{red}(#1,#2)}

\newcommand{\systemc}{SystemC\xspace}

\newcommand{\fairthreads}{FairThreads\xspace}

\newcommand{\osekvdx}{OSEK/VDX\xspace}

\newcommand{\specc}{\textsc{SpecC}\xspace}

\newcommand{\kratos}{\textsc{Kratos}\xspace}

\newcommand{\spin}{\textsc{SPIN}\xspace}
\newcommand{\promela}{\textsc{Promela}\xspace}
\newcommand{\verisoft}{\textsc{VeriSoft}\xspace}
\newcommand{\blast}{\textsc{Blast}\xspace}
\newcommand{\cpachecker}{\textsc{CpaChecker}\xspace}
\newcommand{\satabs}{\textsc{SatAbs}\xspace}

\newcommand{\cbmc}{\textsc{CBMC}\xspace}

\newcommand{\zing}{\textsc{Zing}\xspace}

\newcommand{\ignore}[1]{}
\newcommand{\best}[1]{\textbf{#1}}

%% file: introduction.tex
\section{Introduction}
\label{sec:Intro}


In many practical application domains, the software is organized into
a set of threads that are activated by a scheduler implementing a set
of domain-specific rules. Particularly relevant is the case of
multi-threaded programs with \emph{cooperative scheduling}, \emph{shared-variables} and with
\emph{mutually-exclusive} thread execution. With cooperative scheduling,
there is no preemption: a thread executes, without interruption, until
it either terminates or explicitly yields the control to the
scheduler.
This programming model, simply called \emph{cooperative threads} in
the following, is used in several software paradigms for
embedded systems (e.g.,~
\systemc~\cite{systemc},
\fairthreads~\cite{DBLP:journals/concurrency/Boussinot06},
\osekvdx~\cite{OSEK},
\specc~\cite{GerstlauerEtAl:SystemDesign:01}), and also in other domains
(e.g.,~\cite{DBLP:journals/fac/CimattiGMRTT98}).


Such applications are often critical, and it is thus important to
provide highly effective verification techniques. In this paper, we
consider the use of formal techniques for the verification of
cooperative threads. We face two key difficulties: on the one
side, we must deal with the potentially infinite state space of the
threads, which often requires the use of abstractions; on the other
side, the overall correctness often depends on the details of the
scheduling policy, and thus the use of abstractions in the
verification process may result in false positives.

Unfortunately, the state of the art in verification is unable to deal
with such challenges. Previous attempts to apply various software
model checking techniques to cooperative threads (in specific domains)
have demonstrated limited effectiveness. 
For example, techinques 
like~\cite{DBLP:conf/memocode/KroeningS05,DBLP:conf/spin/TraulsenCMM07,DBLP:journals/fmsd/ClarkeJK07}
abstract away significant aspects of the scheduler and synchronization
primitives, and thus they may report too many false positives, due
to loss of precision, and their applicability is also limited.
Symbolic techniques, 
like~\cite{DBLP:conf/acsd/MoyMM05,DBLP:conf/codes/HerberFG08}, 
show poor scalability because too many details of the scheduler are included 
in the model. Explicit-state techniques, 
like~\cite{DBLP:conf/spin/CampanaCNR11}, are effective in handling the details
of the scheduler and in exploring possible thread interleavings, but
are unable to counter the infinite nature of the state space of 
the threads~\cite{DBLP:journals/sttt/GroceV04}. Unfortunately, for 
explicit-state techniques, a finite-state abstraction is not easily available 
in general.

Another approach could be to reduce the verification of cooperative
threads to the verification of sequential programs.  This approach
relies on a translation from (or \emph{sequentialization} of) the
cooperative threads to the (possibly non-deterministic) sequential
programs that contain both the mapping of the threads in the form of
functions and the encoding of the scheduler. The sequentialized
program can be analyzed by means of ``off-the-shelf'' software model
checking techniques, such as~\cite{DBLP:conf/tacas/ClarkeKSY05,DBLP:conf/cav/McMillan06,DBLP:journals/sttt/BeyerHJM07},
that are based on the counter-example guided abstraction refinement
(\CEGAR)~\cite{DBLP:journals/jacm/ClarkeGJLV03} paradigm.
However, this approach turns out to be problematic. General purpose
analysis techniques are unable to exploit the intrinsic structures of
the combination of scheduler and threads, hidden by the translation into a
single program. For instance, abstraction-based techniques are
inefficient because the abstraction of the scheduler is often too aggressive, and many
refinements are needed to re-introduce necessary details.

In this paper we propose a verification technique which is tailored to the
verification of cooperative threads.  The technique translates each
thread into a separate sequential program; each thread is analyzed, as if it
were a sequential program, with the lazy predicate abstraction
approach~\cite{DBLP:conf/popl/HenzingerJMS02,DBLP:journals/sttt/BeyerHJM07}. 
The overall verification is orchestrated by the direct
execution of the scheduler, with techniques similar to explicit-state
model checking.  This technique, in the following referred to as
\emph{Explicit-Scheduler/Symbolic Threads} (\ESST) model checking,
lifts the lazy predicate abstraction for sequential software to the more
general case of multi-threaded software with cooperative scheduling.

Furthermore, we enhance \ESST with partial-order
reduction~\cite{DBLP:books/sp/Godefroid96,PeledCAV93,ValmariAPN90}. In
fact, despite its relative effectiveness, \ESST often requires the
exploration of a large number of thread interleavings, many of which
are redundant, with subsequent degradations in the run time
performance and high memory
consumption~\cite{DBLP:conf/fmcad/CimattiMNR10}.  \POR essentially
exploits the commutativity of concurrent transitions that result in
the same state when they are executed in different orders.  We
integrate within \ESST two complementary \POR techniques,
\emph{persistent sets} and \emph{sleep sets}. The \POR techniques in
\ESST limit the expansion of the transitions in the explicit
scheduler, while leave the nature of the symbolic analysis of the
threads unchanged.  The integration of \POR in \ESST algorithm is only
seemingly trivial, because \POR could in principle interact negatively
with the lazy predicate abstraction used for analyzing the threads.


The \ESST algorithm has been implemented within the \kratos software model
checker \cite{DBLP:conf/cav/CimattiGMNR11}. \kratos has a generic
structure, encompassing the cooperative threads framework, and has been
specialized for the verification of \systemc programs~\cite{systemc}
and of \fairthreads programs~\cite{DBLP:journals/concurrency/Boussinot06}.
Both \systemc and \fairthreads fall within the paradigm of cooperative
threads, but they have significant differences. This indicates that
the \ESST approach is highly general, and can be adapted to specific
frameworks with moderate effort.
We carried out an extensive experimental evaluation over a significant
set of benchmarks taken and adapted from the literature.  We first
compare \ESST with the verification of sequentialized benchmarks, and
then analyze the impact of partial-order reduction. The results
clearly show that \ESST dramatically outperforms the approach based on
sequentialization, and that both \POR techniques are very effective in
further boosting the performance of \ESST.


This paper presents in a general and coherent manner material
from~\cite{DBLP:conf/fmcad/CimattiMNR10} and
from~\cite{DBLP:conf/tacas/CimattiNR11}.
While in~\cite{DBLP:conf/fmcad/CimattiMNR10}
and in ~\cite{DBLP:conf/tacas/CimattiNR11} the focus is on \systemc,
the framework presented in this paper deals with the general case of
cooperative threads, without focussing on a specific programming
framework.
In order to emphasize the generality of the approach, the
experimental evaluation in this paper has been carried out in
a completely different setting than the one used
in~\cite{DBLP:conf/fmcad/CimattiMNR10}
and in~\cite{DBLP:conf/tacas/CimattiNR11}, namely the \fairthreads
programming framework.
We also considered a set of new benchmarks
from~\cite{DBLP:journals/concurrency/Boussinot06} and
from~\cite{DBLP:journals/rts/WaszniowskiH08}, in addition to adapting
some of the benchmarks used in~\cite{DBLP:conf/tacas/CimattiNR11} to the
\fairthreads scheduling policy.
We also provide proofs of correctness of the proposed techniques in
Appendix~\ref{sec:appendixProofs}.


The structure of this paper is as follows. 
Section~\ref{sec:Background} provides some background in software
model checking via the lazy predicate abstraction.
Section~\ref{sec:ProgModel} introduces the programming model to which
\ESST can be applied.
Section~\ref{sec:ESST} presents the \ESST algorithm.
Section~\ref{sec:ESSTPlusPOR} explains how to extend \ESST with
\POR techniques.
Section~\ref{sec:Application} shows the experimental evaluation.
Section~\ref{sec:RelatedWork} discusses some related work.
Finally, Section~\ref{sec:ConclusionFutureWork} draws conclusions and
outlines some future work.

%% file: background.tex
\section{Background}
\label{sec:Background}

In this section we provide some background on software model checking
via the lazy predicate abstraction for sequential programs.

\subsection{Sequential Programs}
\label{subsec:ProgRep}

We consider sequential programs written in a simple imperative
programming language over a finite set $\Var$ of integer variables, with
basic control-flow constructs (e.g., sequence, if-then-else, iterative
loops) where each \emph{operation}
is either an assignment or an assumption.
An \emph{assignment} is of the form $x := exp$, where $x$ is a
variable and $exp$ is either a variable, an integer constant, an
explicit nondeterministic construct $*$, or an arithmetic operation.
To simplify the presentation, we assume that the considered programs
do not contain function calls. Function calls can be removed by
inlining, under the assumption that there are no recursive calls (a
typical assumption in embedded software).
An \emph{assumption} is of the form $[bexp]$, where $bexp$ is a
Boolean expression that can be a relational operation or an operation
involving Boolean operators.
Subsequently, we denote by $Ops$ the set of program operations.

Without loss of generality, we represent a program $P$ by a
control-flow graph (\CFG).
\begin{defi}[Control-Flow Graph]
\label{def:CFG}
A \emph{control-flow graph} $G$ for a program $P$ is a tuple
$(L,E,l_0,L_{err})$ where 
\begin{enumerate}[(1)]
  \item $L$ is the set of program locations, 
  \item $E \subseteq L \times Ops \times L$ is the set of directed edges
        labelled by a program operation from the set $Ops$,
  \item $l_0 \in L$ is the unique entry location such that,
        for any location $l \in L$ and any operation $op \in Ops$, 
        the set $E$ does not contain any edge $(l,op,l_0)$, and
  \item $L_{err} \subseteq L$ of is the set of \emph{error locations} such
        that, for each $l_e \in L_{err}$, we have $(l_e,op,l) \not\in
        E$ for all $op \in Ops$ and for all $l \in L$.
\end{enumerate}
\end{defi}

\noindent In this paper we are interested in verifying safety
properties by reducing the verification problem to the reachability of
error locations.

\begin{figure}
\begin{scriptsize}
\psset{xunit=0.7cm,yunit=0.7cm,labelsep=2pt}
\vspace*{-5mm}
\begin{pspicture}(0,0)(6,9)
\rput(2,8.5){\circlenode{l0}{$l_0$}}
\rput(2,7.2){\circlenode{l1}{$l_1$}}
\rput(3.5,5.9){\circlenode{l2}{$l_2$}}
\rput(0.5,5.9){\circlenode{l8}{$l_8$}}
\rput(3.5,4.6){\circlenode{l3}{$l_3$}}
\rput(2,3.3){\circlenode{l4}{$l_4$}}
\rput(5,3.3){\circlenode{l5}{$l_5$}}
\rput(3.5,2){\circlenode{l6}{$l_6$}}
\rput(2,0.7){\circlenode[fillstyle=solid,fillcolor=red]{le}{${\color{white}l_e}$}}
\rput(5,0.7){\circlenode{l7}{$l_7$}}
\ncline{->}{l0}{l1}
\ncline{->}{l1}{l2}
\naput{\texttt{[true]}}
\ncline{->}{l1}{l8}
\nbput{\texttt{[false]}}
\ncline{->}{l2}{l3}
\naput{\texttt{x := *}}
\ncline{->}{l3}{l4}
\nbput{\texttt{[x < 0]}}
\ncline{->}{l3}{l5}
\naput{\texttt{[x >= 0]}}
\ncline{->}{l4}{l6}
\nbput{\texttt{y := -x}}
\ncline{->}{l5}{l6}
\naput{\texttt{y := x}}
\ncline{->}{l6}{l7}
\naput{\texttt{[y >= 0]}}
\ncline{->}{l6}{le}
\nbput{\texttt{[y < 0]}}
\ncangles[linearc=.2,armA=.5]{->}{l7}{l1}
\end{pspicture}
\vspace*{-5mm}
\end{scriptsize}
\caption{An example of a\newline control-flow graph.\label{fig:CFG}}
\end{figure}
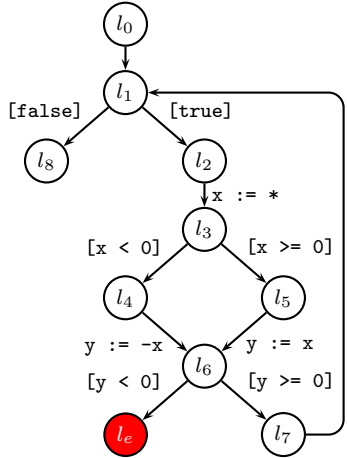
\begin{exa}
\label{ex:CFG}
Figure~\ref{fig:CFG} depicts an example of a \CFG. 
Typical program assertions can be represented by branches going to error 
locations. For example, the branches going out of $l_6$ can be 
the representation of \code{assert(y >= 0)}.
\end{exa}

A \emph{state} $\st$ of a program is a mapping from variables to their
values (in this case integers).  
Let $\State$ be the set of states, we have
$
  \st \in \State = \Var \rightarrow \Val.
$
We denote by $\StDom{\st}$ the domain of a state $\st$. 
We also denote by
$
  \st[x_1 \mapsto v_1, \ldots, x_n \mapsto v_n]
$
the state obtained from $\st$ by substituting the image of $x_i$ in
$\st$ by $v_i$ for all $i = 1, \ldots, n$. 
Let $G = (L,E,l_0,L_{err})$ be the \CFG for a program $P$. 
A \emph{configuration} $\cg$ of $P$ is a pair $(l,\st)$, where $l \in
L$ and $\st$ is a state.
We assume some first-order language in which one can represent a set
of states symbolically. 
We write $\st \models \varphi$ to mean the formula $\varphi$ is true
in the state $\st$, and also say that $\st$ \emph{satisfies}
$\varphi$, or that $\varphi$ \emph{holds at} $\st$.  
A \emph{data region} $\reg \subseteq \State$ is a set of states. 
A data region $\reg$ can be represented symbolically by a first-order
formula $\varphi_{\reg}$, with free variables from $\Var$, such that
all states in $\reg$ satisfy $\varphi_{\reg}$; that is, $\reg =
\setof{\st \mid \st \models \varphi_{\reg}}$.
When the context is clear, we also call the formula $\varphi_{\reg}$ 
data region as well. 
An \emph{atomic region}, or simply a \emph{region}, is a pair
$(l,\varphi)$, where $l \in L$ and $\varphi$ is a data region,
such that the pair represents the set 
$\setof{(l,\st) \mid \st \models \varphi}$ of program configurations.
When the context is clear, we often refer to the both kinds of region
as simply region.

The semantics of an operation $op \in Ops$ can be defined by the
\emph{strongest post-operator} $\SPO{op}$. 
For a formula $\varphi$ representing a region, the \emph{strongest
post-condition} $\SP{\varphi}{op}$ represents the set of states that
are reachable from any of the states in the region represented by
$\varphi$ after the execution of the operation $op$.
The semantics of assignment and assumption operations are as follows:
\[
\begin{array}{rcl}
  \SP{\varphi}{x := exp} & = & \exists x'. \varphi[x/x'] \AND (x = exp[x/x']) 
  \mbox{, for $exp \neq *$,}\\
  \SP{\varphi}{x := *}   & = & \exists x'. \varphi[x/x'] \AND (x = a) \mbox
  {, where $a$ is a fresh variable, and}\\
  \SP{\varphi}{[bexp]}   & = & \varphi \AND bexp,
\end{array}
\]
where $\varphi[x/x']$ and $exp[x/x']$, respectively, denote the
formula obtained from $\varphi$ and the expression obtained from $exp$
by replacing the variable $x'$ for $x$.  
We define the application of the strongest post-operator to a finite sequence 
$\opsseq = op_1,\ldots,op_n$ of operations as the successive application 
of the strongest post-operator to each operator as follows:
$\SP{\varphi}{\opsseq} = \SP{\ldots \SP{\varphi}{op_1} \ldots}{op_n}$.

\subsection{Predicate Abstraction}
\label{subsec:PredAbs}
A program can be viewed as a transition system with transitions between
configurations. The set of configurations can potentially be infinite
because the states can be infinite. 
Predicate abstraction~\cite{DBLP:conf/cav/GrafS97} is a technique 
for extracting a finite transition system from a potentially infinite one 
by approximating possibly infinite sets of states of the latter system 
by Boolean combinations of some predicates.

Let $\Preds$ be a set of predicates over program variables in some
quantifier-free theory $\Theory$. A \emph{precision} $\pi$ is a finite
subset of $\Preds$.
A \emph{predicate abstraction $\varphi^\pi$ of a formula} $\varphi$
over a precision $\pi$ is a Boolean formula over $\pi$ that is
entailed by $\varphi$ in $\Theory$, that is, the formula $\varphi
\IMPL \varphi^\pi$ is valid in $\Theory$. To avoid losing precision,
we are interested in the strongest Boolean combination $\varphi^\pi$,
which is called \emph{Boolean predicate abstraction}~\cite{DBLP:conf/cav/LahiriNO06}.
As described in~\cite{DBLP:conf/cav/LahiriNO06}, for a formula $\varphi$,
the more predicates we have in the precision $\pi$, the more expensive 
the computation of Boolean predicate abstraction. We refer the reader 
to~\cite{DBLP:conf/cav/LahiriNO06,DBLP:conf/fmcad/CavadaCFKRS07,DBLP:conf/fmcad/CimattiDJR09}
for the descriptions of advanced techniques for computing predicate abstractions
based on Satisfiability Modulo Theory (SMT)~\cite{DBLP:series/faia/BarrettSST09}.

Given a precision $\pi$, we can define the \emph{abstract
strongest post-operator} $\ASPO{op}{\pi}$ for an operation $op$. That is,
the \emph{abstract strongest post-condition} $\ASP{\varphi}{op}{\pi}$ is 
the formula $(\SP{\varphi}{op})^\pi$.

\subsection{Predicate-Abstraction based Software Model Checking}
\label{subsec:SWMC}

One prominent software model checking technique is the lazy predicate
abstraction~\cite{DBLP:journals/sttt/BeyerHJM07} technique. 
This technique is a counter-example guided abstraction refinement
(\CEGAR)~\cite{DBLP:journals/jacm/ClarkeGJLV03} technique based on
on-the-fly construction of an \emph{abstract reachability tree}
(\ART). 
An \ART describes the reachable abstract states of the program: a node
in an \ART is a region $(l,\varphi)$ describing an abstract
state. 
Children of an \ART node (or \emph{abstract successors}) are obtained
by unwinding the \CFG and by computing the abstract post-conditions of
the node's data region with respect to the unwound \CFG edge and some
precision $\pi$.
That is, the abstract successors of a node $(l,\varphi)$ is the set
$\setof{(l_1,\varphi_1),\ldots,(l_n,\varphi_n)}$, where, for
$i=1,\ldots,n$, we have 
$(l,op_i,l_i)$ is a \CFG edge, and 
$\varphi_i = \ASP{\varphi}{op_i}{\pi_i}$ for some precision $\pi_i$.  
The precision $\pi_i$ can be associated with the location $l_i$ or can
be associated globally with the \CFG itself. 
The \ART edge connecting a node $(l,\varphi)$ with its child
$(l',\varphi')$ is labelled by the operation $op$ of the \CFG edge
$(l,op,l')$.
In this paper computing abstract successors of an \ART node is also called 
node expansion. 
An \ART node $(l,\varphi)$ is \emph{covered} by another \ART node
$(l',\varphi')$ if $l = l'$ and $\varphi$ entails $\varphi'$.
A node $(l,\varphi)$ can be expanded if it is not covered
by another node and its data region $\varphi$ is satisfiable.
An \ART is \emph{complete} if no further node expansion is possible.
An \ART node $(l,\varphi)$ is an \emph{error node} if $\varphi$ is
satisfiable and $l$ is an error location. 
An \ART is \emph{safe} if it is complete and does not contain any
error node. Obtaining a safe \ART implies that the program is safe.

The construction of an \ART for a the \CFG $G = (L,E,l_0,L_{err})$ for
a program $P$ starts from its root $(l_0,\top)$.
During the construction, when an error node is reached, we check if 
the path from the root to the error node is feasible. 
An \ART path $\tpath$ is a finite sequence $\tedge_1,\ldots,\tedge_n$
of edges in the \ART such that, for every $i=1,\ldots,n-1$, the target
node of $\tedge_i$ is the source node of $\tedge_{i+1}$.  
Note that, the \ART path $\tpath$ corresponds to a path in the \CFG.
We denote by $\pathops{\tpath}$ the sequence of operations labelling
the edges of the \ART path $\tpath$. 
A counter-example path is an \ART path $\tedge_1,\ldots,\tedge_n$ such
that the source node of $\tedge_1$ is the root of the \ART and the
target node of $\tedge_n$ is an error node.  
A counter-example path $\tpath$ is \emph{feasible} if and only if
$\SP{true}{\pathops{\tpath}}$ is satisfiable. An infeasible
counter-example path is also called \emph{spurious} counter-example.
A feasible counter-example path witnesses that the program $P$ is
unsafe.

An alternative way of checking feasibility of a counter-example path
$\tpath$ is to create a \emph{path formula} that corresponds to the
path.
This is achieved by first transforming the sequence $\pathops{\tpath}
= op_1,\ldots,op_n$ of operations labelling $\tpath$ into its
single-static assignment (SSA)
form~\cite{DBLP:journals/toplas/CytronFRWZ91}, where there is only one
single assignment to each variable.
Next, a constraint for each operation is generated by rewriting
each assignment $x := exp$ into the equality $x = exp$, with nondeterministic
construct $*$ being translated into a fresh variable, and turning
each assumption $[bexp]$ into the constraint $bexp$. 
The path formula is the conjunction of the constraint generated by 
each operation.  
A counter-example path $\tpath$ is feasible if and only if its 
corresponding path formula is satisfiable.

\begin{exa}
Suppose that the operations labelling a counter-example path are
\[
  \mathtt{x := y}, 
  \ \mathtt{[x > 0]}, 
  \ \mathtt{x := x + 1}, 
  \ \mathtt{y := x},
  \ \mathtt{[y < 0]},
\]
then, to check the feasibility of the path, we check the satisfiability
of the following formula:
\[
  \mathtt{x_1 = y_0}
  \AND \mathtt{x_1 > 0}
  \AND \mathtt{x_2 = x_1 + 1}
  \AND \mathtt{y_1 = x_2}
  \AND \mathtt{y_1 < 0}.
\]
\end{exa} 

If the counter-example path is infeasible, then it has to be removed 
from the constructed \ART by refining the precisions. Such a refinement
amounts to analyzing the path and extracting new predicates from it.
One successful method for extracting relevant predicates at
certain locations of the \CFG is based on the computation of Craig
interpolants~\cite{Craig:JSymLogic:1957}, as shown
in~\cite{DBLP:conf/popl/HenzingerJMM04}.
Given a pair of formulas $(\varphi^-,\varphi^+)$ such that $\varphi^-
\AND \varphi^+$ is unsatisfiable, a \emph{Craig interpolant} of
$(\varphi^-,\varphi^+)$ is a formula $\psi$ such that $\varphi^-
\IMPL \psi$ is valid, $\psi \AND \varphi^+$ is unsatisfiable, and
$\psi$ contains only variables that are common to both $\varphi^-$
and $\varphi^+$.
Given an infeasible counter-example $\tpath$, the predicates can be
extracted from interpolants in the following way:
\begin{enumerate}[(1)]
  \item Let $\pathops{\tpath} = op_1,\ldots,op_n$, and let 
  the sub-path $\pathops{\tpath}^{i,j}$ such that $i \leq j$ denote
  the sub-sequence $op_i,op_{i+1},\ldots,op_j$ of $\pathops{\tpath}$.
  
  \item For every $k = 1,\ldots,n-1$, let $\varphi^{1,k}$ be the path
  formula for the sub-path  $\pathops{\tpath}^{1,k}$ and  
  $\varphi^{k+1,n}$ be the path formula for the sub-path 
  $\pathops{\tpath}^{k+1,n}$, we generate an interpolant 
  $\psi^k$ of $(\varphi^{1,k},\varphi^{k+1,n})$.

  \item The predicates are the (un-SSA) atoms in the interpolant $\psi^k$ for
  $k = 1,\ldots,n$.
\end{enumerate}
The discovered predicates are then added to the precisions that are
associated with some locations in the \CFG. 
Let $p$ be a predicate extracted from the interpolant $\psi^k$ of
$(\varphi^{1,k},\varphi^{k+1,n})$ for $1 \leq k < n$. 
Let $\tedge_1,\ldots,\tedge_n$ be the sequence of edges labelled by
the operations $op_1,\ldots,op_n$, that is, for $i=1,\ldots,n$, the
edge $\tedge_i$ is labelled by $op_i$. 
Let the nodes $(l,\varphi)$ and $(l',\varphi')$ be the source and
target nodes of the edge $\tedge_k$. 
The predicate $p$ can be added to the precision associated with the
location $l'$.

Once the precisions have been refined, the constructed \ART is
analyzed to remove the sub part containing the infeasible
counter-example path, and then the \ART is reconstructed using the refined
precisions.

Lazy predicate abstraction has been implemented in several software
model checkers, including \blast~\cite{DBLP:journals/sttt/BeyerHJM07},
\cpachecker~\cite{DBLP:conf/cav/BeyerK11}, and
\kratos~\cite{DBLP:conf/cav/CimattiGMNR11}.  
For details and in-depth illustrations of \ART constructions, we refer
the reader to~\cite{DBLP:journals/sttt/BeyerHJM07}.

%% file: program-model.tex
\section{Programming Model}
\label{sec:ProgModel}

In this paper we analyze shared-variable multi-threaded programs with
\emph{exclusive thread} (there is at most one running thread at a
time) and \emph{cooperative scheduling policy} (the scheduler never
preempts the running thread, but waits until the running thread
cooperatively yields the control back to the scheduler).
At the moment we do not deal with dynamic thread creations. This
restriction is not severe because typically multi-threaded programs
for embedded system designs are such that all threads are known and
created a priori, and there are no dynamic thread creations.

\begin{figure}
\centerline{\includegraphics[scale=0.8]{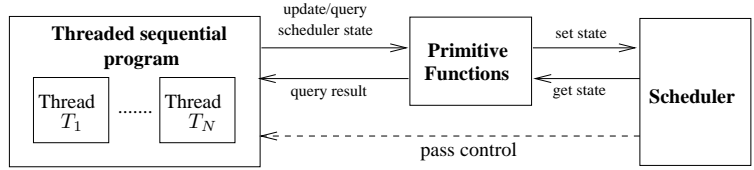}}
\caption{Programming model.\label{fig:ProgModel}} 
\end{figure}

Our programming model is depicted in Figure~\ref{fig:ProgModel}.  
It consists of three components: a so-called threaded sequential
program, a scheduler, and a set of primitive functions.  
A \emph{threaded sequential program} (or \emph{threaded program}) $P$
is a multi-threaded program consisting of a set of sequential programs
$T_1,\ldots,T_N$ such that each sequential program $T_i$ represent a
\emph{thread}. From now on, we will refer to the sequential programs
in the threaded programs as threads.
We assume that the threaded program has a main thread, denoted
by $\mainthread$, from which the execution starts. The main thread is 
responsible for initializing the shared variables.

Let $P$ be a threaded program, we denote by $\GVar$ the set of shared
(or global) variables of $P$ and by $\TLVar{T}$ the set of local
variables of the thread $T$ in $P$.
We assume that $\TLVar{T} \cap \GVar = \emptyset$ for every
thread $T$ and $\TLVar{T_i} \cap \TLVar{T_j} = \emptyset$ 
for each two threads $T_i$ and $T_j$ such that $i \neq j$.
We denote by $\TCFG{T}$ the \CFG for the thread $T$. All operations in
$\TCFG{T}$ only access variables in $\TLVar{T} \cup \GVar$.

The \emph{scheduler} governs the executions of threads. 
It employs a cooperative scheduling policy that only allows at most
one running thread at a time.  
The scheduler keeps track of a set of variables that are necessary to
orchestrate the thread executions and synchronizations. We denote such
a set by $\SVar$. 
For example, the scheduler can keep track of the states of threads and
events, and also the time delays of event notifications. 
The mapping from variables in $\SVar$ to their values form a
\emph{scheduler state}. 
Passing the control to a thread can be done, for example, by simply
setting the state of the thread to running. 
Such a control passing is represented by the dashed line in
Figure~\ref{fig:ProgModel}.

\emph{Primitive functions} are special functions used by the threads
to communicate with the scheduler by querying or updating the
scheduler state.
To allow threads to call primitive functions, we simply extend the
form of assignment described in Section~\ref{subsec:ProgRep} as
follows: the expression $exp$ of an assignment $x := exp$ can also be
a call to a primitive function.
We assume that such a function call is the top-level expression $exp$
and not nested in another expression. 
Calls to primitive functions do not modify the values of variables
occurring in the threaded program. 
Note that, as primitive function calls only occur on the right-hand
side of assignment, we implicitly assume that every primitive function
has a return value.

The primitive functions can be thought of as a programming interface
between the threads and the scheduler.
For example, for event-based synchronizations, one can have a
primitive function \code{wait\_event($e$)} that is parametrized by an
event name $e$.
This function suspends the calling thread by telling the scheduler
that it is now waiting for the notification of event $e$.
Another example is the function \code{notify\_event($e$)} that triggers 
the notification of event $e$ by updating the event's state, which is 
tracked by the scheduler, to a value indicating that it has been notified.
In turn, the scheduler can wake up the threads that are waiting for the
notification of $e$ by making them runnable.

We now provide a formal semantics for our programming model.
Evaluating expressions in program operations involves three kinds of state:
\begin{enumerate}[(1)]
  \item The state $\st_i$ of local variables of some thread $T_i$ 
        ($\StDom{\st_i} = \TLVar{T_i}$).
  \item The state $\gst$ of global variables ($\StDom{\gst} = \GVar$).
  \item The \emph{scheduler state} $\sst$ ($\StDom{\sst} = \SVar$). 
\end{enumerate}
The evaluation of the right-hand side expression of an assignment
requires a scheduler state because the expression can be a call to a
primitive function whose evaluation depends on and can update the
scheduler state.

We require, for each thread $T$, there is a variable $\tstatus{T} \in
\StDom{\sst}$ that indicates the state of $T$.
We consider the set $\setof{\RUNNING, \RUNNABLE, \WAITING}$ as the
domain of $\tstatus{T}$, where each element in the set has an obvious
meaning.
The elements $\RUNNING$, $\RUNNABLE$, and $\WAITING$ can be thought of
as enumerations that denote different integers. 
We say that the thread $T$ is \emph{running}, \emph{runnable}, or
\emph{waiting} in a scheduler state $\sst$ if $\sst(\tstatus{T})$ is,
respectively, $\RUNNING$, $\RUNNABLE$, or $\WAITING$. 
We denote by $\schedstates$ the set of all scheduler states. 
Given a threaded program with $N$ threads $T_1,\ldots,T_N$, by the exclusive 
running thread property, we have, for every state $\sst \in \schedstates$, 
if, for some $i$, we have $\sst(\tstatus{T_i}) = \RUNNING$, then 
$\sst(\tstatus{T_j}) \neq \RUNNING$ for all $j \neq i$, where 
$1 \leq i,j \leq N$.

The semantics of expressions in program operations are given by the
following two evaluation functions
\[
\begin{array}{lcl}
  \semE{\cdot} & : & exp 
                     \rightarrow 
                     ((\State \times \State \times \schedstates) 
                       \rightarrow (\Val \times \schedstates))  \\

  \semB{\cdot} & : & bexp 
                     \rightarrow 
                     ((\State \times \State \times \schedstates) 
                       \rightarrow \setof{true,false}). \\
\end{array}
\]
The function $\semE{\cdot}$ takes as arguments an expression occurring
on the right-hand side of an assignment and the above three kinds of
state, and returns the value of evaluating the expression over the
states along with the possible updated scheduler state.
The function $\semB{\cdot}$ takes as arguments a boolean expression
and the local and global states, and returns the valuation of the
boolean expression.
Figure~\ref{fig:SemanticsEvalFun} shows the semantics of expressions
in program operations given by the evaluation functions $\semE{\cdot}$
and $\semB{\cdot}$. 
To extract the result of evaluation function, we use the standard
projection function $\proj{i}$ to get the $i$-th value of a tuple. 
The rules for unary arithmetic operations and unary boolean operations
can be defined similarly to their binary counterparts.
\begin{figure}
\begin{tabular}{|p{0.22\textwidth}|p{0.73\textwidth}|}\hline
\small \textbf{Variable} & 
\small $\semE{x}(\st,\gst,\sst) = (v,\sst)$, where $v = \st(x)$ if $x \in
\StDom{\st}$ or $v = \gst(x)$ if $x \in \StDom{\gst}$.\\\hline
\small \textbf{Integer constant} &
\small $\semE{c}(\st,\gst,\sst) = (c,\sst)$.\\\hline
\small \textbf{Nondeterministic\newline construct} &
$\semE{*}(\st,\gst,\sst) = (v,\sst)$, for some $v \in \Val$. \\\hline
\small \textbf{Binary arithmetic\newline operation} &
\small $\semE{exp_1 \otimes exp_2}(\st,\gst,\sst) = (v1 \otimes v2,\sst)$, where 
$v1 = \proj{1}(\semE{exp_1}(\st,\gst,\sst))$ and 
$v2 = \proj{1}(\semE{exp_2}(\st,\gst,\sst))$.\\\hline
\small \textbf{Primitive \newline function call} &
\small $\sem{f(exp_1,\ldots,exp_n)}(\st,\gst,\sst) = (v,\sst')$, where
$(v,\sst') = f'(v_1,\ldots,v_n,\sst)$ and 
$v_i = \proj{1}{\semE{exp_i}(\st,\gst,\sst)}$, for $i = 1,\ldots,n$.\\\hline
\small \textbf{Relational operation} &
\small $\semB{exp_1 \odot exp_2}(\st,\gst,\sst) = v1 \odot v2$, where
$v1 = \proj{1}(\semE{exp_1}(\st,\gst,\sst))$ and 
$v2 = \proj{1}(\semE{exp_2}(\st,\gst,\sst))$.\\\hline
\small \textbf{Binary boolean \newline operation} &
\small $\semB{bexp_1 \star  bexp_2}(\st,\gst,\sst) = v1 \star v2$, where
$v1 = \semB{bexp_1}(\st,\gst,\sst)$ and 
$v2 = \semB{bexp_2}(\st,\gst,\sst)$.\\\hline
\end{tabular}

\caption{Semantics of expressions in program operations. 
\label{fig:SemanticsEvalFun}}
\end{figure}
For primitive functions, we assume that every $n$-ary primitive
function $f$ is associated with an $(n+1)$-ary function $f'$ such that
the first $n$ arguments of $f'$ are the values resulting from the
evaluations of the arguments of $f$, and the $(n+1)$-th argument of 
$f'$ is a scheduler state. 
The function $f'$ returns a pair of value and updated scheduler state.
 
Next, we define the meaning of a threaded program by using the
operational semantics in terms of the \CFGs of the threads.
The main ingredient of the semantics is the notion of run-time configuration. 
Let $\TCFG{T} = (L,E,l_0,L_{err})$ be the \CFG for a thread $T$. A
\emph{thread configuration} $\tcg{T}$ of $T$ is a pair $(l,\st)$,
where $l \in L$ and $\st$ is a state such that $\StDom{\st} =
\TLVar{T}$.

\begin{defi}[Configuration]\label{def:Config}
A \emph{configuration} $\cg$ of a threaded program $P$ with $N$
threads $T_1,\ldots,T_N$ is a tuple
$\conf{\tcg{T_1},\ldots,\tcg{T_N}}{\gst}{\sst}$ where
\begin{iteMize}{$\bullet$}
  \item each $\tcg{T_i}$ is a thread configuration of thread $T_i$,
  \item $\gst$ is the state of global variables, and
  \item $\sst$ is the scheduler state.
\end{iteMize}
\end{defi}
For succinctness, we often refer the thread configuration $\tcg{T_i} =
(l,\st)$ of the thread $T_i$ as the indexed pair $(l,\st)_i$.
A configuration $\conf{\tcg{T_1},\ldots,\tcg{T_N}}{\gst}{\sst}$, is an
\emph{initial configuration} for a threaded program if
for each $i=1,\ldots,N$, the location $l$ of $\tcg{T_i}=(l,\st)$ is
the entry of the \CFG $\TCFG{T_i}$ of $T_i$, and 
$\sst(\tstatus{\mainthread}) = \RUNNING$ and $\sst(\tstatus{T_i}) \neq
\RUNNING$ for all $T_i \neq \mainthread$.

Let $\nrschedstates \subset \schedstates$ be the set of scheduler
states such that every state in $\nrschedstates$ has no running
thread, and $\orschedstates \subset \schedstates$ be the set of
scheduler states such that every state in $\orschedstates$ has exactly
one running thread. 
A scheduler with a cooperative scheduling policy can simply be
defined as a function $\SchedFn : \nrschedstates \rightarrow
\powerset{\orschedstates}$. 

The transitions of the semantics are of the form
\[
\begin{array}{ll}
  \mbox{Edge transition: }     & \cg \transsem{op} \cg' \\
  \mbox{Scheduler transition:} & \cg \transsem{\cdot} \cg' 
\end{array}
\]
where $\cg,\cg'$ are configurations and $op$ is the operation labelling 
an edge.
Figure~\ref{fig:OpSem} shows the semantics of threaded programs.
The first three rules show that transitions over edges of 
the \CFG $\TCFG{T}$ of a thread $T$ are defined if and 
only if $T$ is running, as indicated by the scheduler state. 
The first rule shows that a transition over an edge labelled by an
assumption is defined if the boolean expression of the assumption 
evaluates to true.  
The second and third rules show the updates of the states caused by
the assignment.
Finally, the fourth rule describes the running of the scheduler.

\begin{figure}
\begin{tabular}{c}
\begin{small}
$
\infer[\hspace*{2mm}(1)]
{ \conf{\tcg{T_1},\ldots,(l,\st)_i,\ldots,\tcg{T_N}}{\gst}{\sst} 
  \transsem{[bexp]} 
  \conf{\tcg{T_1},\ldots,(l',\st)_i,\ldots,\tcg{T_N}}{\gst}{\sst} }
{ \TCFG{T_i} = (L,E,l_0,L_{err}) 
  & (l,[bexp],l') \in E
  & \sst(\tstatus{T_i}) = \RUNNING 
  & \semB{[bexp]}(\st,\gst,\sst) = true }
$
\end{small}\\[6mm]
\begin{small}
$
\infer[\hspace*{2mm}(2)]
{ \conf{\tcg{T_1},\ldots,(l,\st)_i,\ldots,\tcg{T_N}}{\gst}{\sst} 
  \transsem{x := exp} 
  \conf{\tcg{T_1},\ldots,(l',\st')_i,\ldots,\tcg{T_N}}{\gst}{\sst'} }
{ \begin{array}{c}
    \TCFG{T_i} = (L,E,l_0,L_{err}) \\
    \semE{x := exp}(\st,\gst,\sst) = (v,\sst') 
  \end{array}
  &
  \begin{array}{c}
    (l,x := exp,l') \in E \\
    \st' = \st[x \mapsto v]
  \end{array}
  &
  \begin{array}{c}
    \sst(\tstatus{T_i}) = \RUNNING \\
     x \in \TLVar{T_i}
  \end{array} }
$
\end{small}\\[6mm]
\begin{small}
$
\infer[\hspace*{2mm}(3)]
{ \conf{\tcg{T_1},\ldots,(l,\st)_i,\ldots,\tcg{T_N}}{\gst}{\sst} 
  \transsem{x := exp} 
  \conf{\tcg{T_1},\ldots,(l',\st)_i,\ldots,\tcg{T_N}}{\gst'}{\sst'} }
{ \begin{array}{c}
    \TCFG{T_i} = (L,E,l_0,L_{err}) \\
    \semE{x := exp}(\st,\gst,\sst) = (v,\sst') 
  \end{array}
  &
  \begin{array}{c}
    (l,x := exp,l') \in E \\
    \gst' = \gst[x \mapsto v]
  \end{array}
  &
  \begin{array}{c}
    \sst(\tstatus{T_i}) = \RUNNING \\
     x \in \GVar
  \end{array} }
$
\end{small}\\[6mm]
\begin{small}
$
\infer[\hspace*{2mm}(4)]
{ \conf{\tcg{T_1},\ldots,\tcg{T_N}}{\gst}{\sst} 
  \transsem{\cdot} 
  \conf{\tcg{T_1},\ldots,\tcg{T_N}}{\gst}{\sst'} }
{ \forall i. \sst(\tstatus{T_i}) \neq \RUNNING
  & \sst' \in \SchedFn(\sst) }
$
\end{small}
\end{tabular}
\caption{Operational semantics of threaded sequential programs.
\label{fig:OpSem}}
\end{figure}

\begin{defi}[Computation Sequence, Run, Reachable Configuration]
\label{def:CompSeq}

A \emph{computation sequence} $\cg_0,\cg_1,\ldots$ of a threaded
program $P$ is either a finite or an infinite sequence of
configurations of $P$ such that, for all $i$, either $\cg_i
\transsem{op} \cg_{i+1}$ for some operation $op$ or $\cg_i
\transsem{\cdot} \cg_{i+1}$. 
A \emph{run} of a threaded program $P$ is a computation sequence
$\cg_0,\cg_1,\ldots$ such that $\cg_0$ is an initial configuration. 
A configuration $\cg$ of $P$ is \emph{reachable from a configuration}
$\cg'$ if there is a computation sequence $\cg_0,\ldots,\cg_n$ such
that $\cg_0 = \cg'$ and $\cg_n = \cg$. 
A configuration $\cg$ is \emph{reachable in $P$} if it is reachable
from an initial configuration.
\end{defi}

A configuration $
\conf{\tcg{T_1},\ldots,(l,\st)_i,\ldots,\tcg{T_N}}{\gst}{\sst} $ of a
threaded program $P$ is an \emph{error configuration} if \CFG
$\TCFG{T_i} = (L,E,l_0,L_{err})$ and $l \in L_{err}$.
We say a threaded program $P$ is \emph{safe} iff no error
configuration is reachable in $P$; otherwise, $P$ is unsafe.

%% file: esst.tex
\section{Explicit-Scheduler Symbolic-Thread (\ESST)}
\label{sec:ESST}

In this section we present our novel technique for verifying threaded
programs. We call our technique \emph{Explicit-Scheduler
Symbolic-Thread} (\ESST)~\cite{DBLP:conf/fmcad/CimattiMNR10}.
This technique is a \CEGAR based technique that combines
explicit-state techniques with the lazy predicate abstraction 
described in Section~\ref{subsec:SWMC}.
In the same way as the lazy predicate abstraction, \ESST analyzes the
data path of the threads by means of predicate abstraction and
analyzes the flow of control of each thread with explicit-state
techniques.  
Additionally, \ESST includes the scheduler as part of its model
checking algorithm and analyzes the state of the scheduler with
explicit-state techniques.

\subsection{Abstract Reachability Forest (\ARF)}
\label{subsec:ARF}

The \ESST technique is based on the on-the-fly construction and
analysis of an \emph{abstract reachability forest} (\ARF). 
An \ARF describes the reachable abstract states of the threaded
program. 
It consists of connected \emph{abstract reachability trees} (\ARTs),
each describing the reachable abstract states of the running thread.
The connections between one \ART with the others in an \ARF describe
possible thread interleavings from the currently running thread to the
next running thread.

Let $P$ be a threaded program with $N$ threads $T_1,\ldots,T_N$.
A \emph{thread region for the thread $T_i$}, for $1 \leq i \leq N$, is 
a set of thread configurations such that the domain of the states of
the configurations is $\TLVar{T_i} \cup \GVar$.
A \emph{global region for a threaded program $P$} is a set of states
whose domain is $\bigcup_{i=1,\ldots,N}\TLVar{T_i} \cup \GVar$.

\begin{defi}[\ARF Node]
An \emph{\ARF node} for a threaded program $P$ with $N$ threads
$T_1,\ldots,T_N$ is a tuple
\[
  (\ststate{l_1}{\varphi_1},\ldots,\ststate{l_N}{\varphi_N},\varphi,\sst), 
\]
where 
$(l_i,\varphi_i)$, for $i=1,\ldots,N$, is a thread region for 
$T_i$, 
$\varphi$ is a global region, and 
$\sst$ is the scheduler state.
\end{defi}

Note that, by definition, the global region, along with the program
locations and the scheduler state, is sufficient for representing the
abstract state of a threaded program. 
However, such a representation will incur some inefficiencies in
computing the predicate abstraction.  
That is, without any thread regions, the precision is only associated
with the global region. 
Such a precision will undoubtedly contains a lot of predicates about
the variables occurring in the threaded program. 
However, when we are interested in computing an abstraction of a
thread region, we often do not need the predicates consisting only 
of variables that are local to some other threads.

In \ESST  we can associate a precision with 
a location $l_i$ of the \CFG $\TCFG{T}$ for thread $T$, denoted by 
$\lprec{l_i}$, 
with a thread $T$, denoted by $\tprec{T}$, or
the global region $\varphi$, denoted by $\gprec$.
For a precision $\tprec{T}$ and for every location $l$ of $\TCFG{T}$, 
we have $\tprec{T} \subseteq \lprec{l}$ for the precision $\lprec{l}$ 
associated with the location $l$.
Given a predicate $\psi$ and a location $l$ of the \CFG $\TCFG{T_i}$, 
and let $\fvar{\psi}$ be the set of free variables of $\psi$, 
we can add $\psi$ into the following precisions:
\begin{iteMize}{$\bullet$}
\item If $\fvar{\psi} \subseteq \TLVar{T_i}$, then 
      $\psi$ can be added into $\gprec$, $\tprec{T_i}$, or $\lprec{l}$.

\item If $\fvar{\psi} \subseteq \TLVar{T_i} \cup \GVar$, then 
      $\psi$ can be added into $\gprec$, $\tprec{T_i}$, or $\lprec{l}$.
\item If $\fvar{\psi} \subseteq \bigcup_{j=1,\ldots,N} 
      \TLVar{T_j} \cup \GVar$, then $\psi$ can be added into $\gprec$.
\end{iteMize}

\subsection{Primitive Executor and Scheduler}
\label{subsec:PrimExecScheduler}

As indicated by the operational semantics of threaded programs,
besides computing abstract post-conditions, we need to execute calls
to primitive functions and to explore all possible schedules (or
interleavings) during the construction of an \ARF.
For the calls to primitive functions, we assume that the values passed
as arguments to the primitive functions are known statically.
This is a limitation of the current \ESST algorithm, and we will
address this limitation in our future work.

Recall that, $\schedstates$ denotes the set of scheduler states, and
let $\primcalls$ be the set of calls to primitive functions.
To implement the semantic function $\semE{exp}$, where $exp$ is a
primitive function call, we introduce the function
\[ 
  \Sexec : (\schedstates \times \primcalls) 
           \rightarrow (\Val \times \schedstates).
\]
This function takes as inputs a scheduler state, a call $f(\vec{x})$
to a primitive function $f$, and returns a value and an updated
scheduler state resulting from the execution of $f$ on the arguments
$\vec{x}$. That is, $\Sexec(\sst,f(\vec{x}))$ essentially
computes $\semE{f(\vec{x})}(\cdot,\cdot,\sst)$.
Since we assume that the values of $\vec{x}$ are known statically, 
we deliberately ignore, by $\cdot$, the states of local and global
variables.

\begin{exa} 
Let us consider a primitive function call \code{wait\_event($e$)} that
suspends a running thread $T$ and makes the thread wait for a
notification of an event $e$. 
Let $\tev{T}$ be the variable in the scheduler state that keeps track
of the event whose notification is waited for by $T$.
The state $\sst'$ of $(\cdot,\sst') =
\Sexec(\sst,\code{wait\_event(e)})$ is obtained from the state $S$ by
changing the status of running thread to $\WAITING$, and noting that
the thread is waiting for event $e$, that is,
$
  \sst' = \sst[\tst{T} \mapsto \WAITING, \tev{T} \mapsto e].
$
\end{exa}

Finally, to implement the scheduler function $\SchedFn$ in the
operational semantics, and to explore all possible schedules, we
introduce the function
\[
  \Sched : \nrschedstates \rightarrow \powerset{\orschedstates}.
\]
This function takes as an input a scheduler state and returns a set of
scheduler states that represent all possible schedules.

\subsection{\ARF Construction}
\label{subsec:ARFConstruct}

We expand an \ARF node by unwinding the \CFG of the running thread and
by running the scheduler. Given an \ARF node
\[ 
  (\ststate{l_1}{\varphi_1},\ldots,\ststate{l_N}{\varphi_N},\varphi,\sst),
\]
we expand the node by the following rules~\cite{DBLP:conf/fmcad/CimattiMNR10}:
\begin{enumerate}[E1.]
\settowidth{\labelwidth}{{E1.}}
\setlength{\leftmargin}{\labelwidth}
\advance \itemindent by 0em
\renewcommand{\theenumi}{{{E}\arabic{enumi}}}
\renewcommand{\makelabel}[1]{\textrm{\theenumi}.}

\item If there is a running thread $T_i$ in $\sst$ such that the thread 
      performs an operation $op$ and $(l_i,op,l'_i)$ is an edge of 
      the \CFG $\TCFG{T_i}$ of thread $T_i$, then we have 
      two cases\label{rule:ARFExp1}:

  \begin{iteMize}{$\bullet$}
  
  \item If $op$ is \emph{not} a call to primitive function, then the successor
        node is 
	\[
	  ( \ststate{l_1}{\varphi'_1},
            \ldots,
	    \ststate{l'_i}{\varphi'_i},
	    \ldots,
            \ststate{l_N}{\varphi'_N},
            \varphi',\sst
          ), 
	\]
	where
        \begin{enumerate}
          \item $\varphi'_i = \ASP{\varphi_i \AND \varphi}{op}{\lprec{l'_i}}$ 
          and $\lprec{l'_i}$ is the precision associated with $l'_i$,
          \item $\varphi_j' = 
              \ASP{\varphi_j \AND \varphi}{\havoc(op)}{\lprec{l_j}}$
          for $j \neq i$ and $\lprec{l_j}$ is the precision associated with 
          $l_j$, if $op$ possibly updates global variables, otherwise
          $\varphi_j' = \varphi_j$, and 
          \item $\varphi' = \ASP{\varphi}{op}{\gprec}$ and $\gprec$ is 
          the precision associated with the global region.
	\end{enumerate}
        The function \havoc collects all global variables possibly
        updated by $op$, and builds a new operation where these
        variables are assigned with fresh variables.
	The edge connecting the original node and the resulting successor
	node is labelled by the operation $op$.

  \item If $op$ is a primitive function call $x := f(\vec{y})$, then 
        the successor node is 
	\[
	  ( \ststate{l_1}{\varphi'_1},
            \ldots,
	    \ststate{l'_i}{\varphi'_i},
	    \ldots,
            \ststate{l_N}{\varphi'_N},
            \varphi',\sst'
          ), 
	\]
	where
        \begin{enumerate}
          \item $(v,\sst') = \Sexec(\sst,f(\vec{y}))$, 
          \item $op'$ is the assignment $x := v$,
          \item $\varphi'_i = \ASP{\varphi_i \AND \varphi}{op'}{\lprec{l'_i}}$
                and $\lprec{l'_i}$ is the precision associated with $l'_i$,
          \item $\varphi_j' = \ASP{\varphi_j \AND
                \varphi}{\havoc(op')}{\lprec{l_j}}$ for $j \neq i$
                and $\lprec{l_j}$ is the precision associated with
                $l_j$ if $op$ possibly updates global variables,
                otherwise $\varphi_j' = \varphi_j$, and
          \item $\varphi' = \ASP{\varphi}{op'}{\gprec}$
                and $\gprec$ is the precision associated with the global region.
	\end{enumerate}
	The edge connecting the original node and the resulting successor
	node is labelled by the operation $op'$.

  \end{iteMize}

\item If there is no running thread in $\sst$, then, for 
      each $\sst' \in \Sched(\sst)$, we create a successor node
      	\[
	  ( \ststate{l_1}{\varphi_1},
            \ldots,
            \ststate{l_N}{\varphi_N},
            \varphi,\sst'
          ).
	\]
      We call such a connection between two nodes an \emph{\ARF connector}.
      \label{rule:ARFExp2}
\end{enumerate}

\noindent Note that, the rule~\ref{rule:ARFExp1} constructs the \ART that
belongs to the running thread, while the connections between the \ARTs
that are established by \ARF connectors in the rule~\ref{rule:ARFExp2}
represent possible thread interleavings or context switches.

An \ARF node  
$ 
  (\ststate{l_1}{\varphi_1},\ldots,\ststate{l_N}{\varphi_N},\varphi,\sst)
$
is the \emph{initial node} if 
for all $i=1,\ldots,N$, the location $l_i$ is the entry location of
the \CFG $\TCFG{T_i}$ of thread $T_i$ and $\varphi_i$ is $true$,
$\varphi$ is $true$, and 
$\sst(\tst{\mainthread}) = \RUNNING$ and $\sst(\tst{T_i}) \neq
\RUNNING$ for all $T_i \neq \mainthread$.

We construct an \ARF by applying the rules~\ref{rule:ARFExp1} and
\ref{rule:ARFExp2} starting from the initial node.
A node can be expanded if the node is not covered by other
nodes and if the conjunction of all its thread regions and the global
region is satisfiable.

\begin{defi}[Node Coverage]\label{def:NodeCoverage}
An \ARF node 
$
  (\ststate{l_1}{\varphi_1},\ldots,\ststate{l_N}{\varphi_N},\varphi,\sst)
$ 
is \emph{covered} by another \ARF node
$
  (\ststate{l'_1}{\varphi'_1},\ldots,\ststate{l'_N}{\varphi'_N},\varphi',\sst')
$
if $l_i = l'_i$ for $i = 1,\ldots,N$, 
$\sst = \sst'$, and 
$\varphi \IMPL \varphi'$ and $\bigwedge_{i = 1,\ldots,N} (\varphi_i
\IMPL \varphi'_i)$ are valid.
\end{defi}

An \ARF is \emph{complete} if it is closed under the expansion of
rules~\ref{rule:ARFExp1} and ~\ref{rule:ARFExp2}. An \ARF node
$
  (\ststate{l_1}{\varphi_1},\ldots,\ststate{l_N}{\varphi_N},\varphi,\sst)
$
is an \emph{error node} if $\varphi \AND \bigwedge_{i=1,\ldots,N}
\varphi_i$ is satisfiable, and at least one of the locations
$l_1,\ldots,l_N$ is an error location. 
An \ARF is \emph{safe} if it is complete and does not contain any
error node.

\subsection{Counter-example Analysis}
\label{subsec:CexAnalysis}

Similar to the lazy predicate abstraction for sequential programs, 
during the construction of an \ARF, when we reach an error node, we check 
if the path in the \ARF from the initial node to the error node is feasible.

\begin{defi}[\ARF Path]\label{def:ARFPath}
An \emph{\ARF path} $\fpath =
\tpath_1,\fconn_1,\tpath_2,\ldots,\fconn_{n-1},\tpath_n$ is a finite
sequence of \ART paths $\tpath_i$ connected by \ARF connectors
$\fconn_j$, such that
\begin{enumerate}[(1)]
  \item $\tpath_i$, for $i = 1,\ldots,n$, is an \ART path,
  \item $\fconn_j$, for $j = 1,\ldots,n-1$, is an \ARF connector, and
  \item for every $j = 1, \ldots, n-1$, such that $\tpath_j =
        \tedge^j_1,\ldots,\tedge^j_m$ and $\tpath_{j+1} =
        \tedge^{j+1}_1,\ldots,\tedge^{j+1}_l$, the target node of $\tedge^j_m$
        is the source node of $\fconn_j$ and the source node of
        $\tedge^{j+1}_1$ is the target node of $\fconn_j$.
\end{enumerate}
A \emph{suppressed \ARF path} $\suppath{\fpath}$ of $\fpath$ is the sequence
$
   \tpath_1,\ldots,\tpath_n.
$
\end{defi}

A \emph{counter-example path} $\fpath$ is an \ARF path such that the
source node of $\tedge_1$ of $\tpath_1=\tedge_1,\ldots,\tedge_m$ is
the initial node, and the target node of $\tedge'_k$ of $\tpath_n =
\tedge'_1,\ldots,\tedge'_k$ is an error node.  
Let $\pathops{\suppath{\fpath}}$ denote the sequence of operations labelling 
the edges in $\suppath{\fpath}$.
We say that a counter-example path $\fpath$ is \emph{feasible} if and
only if $\SP{true}{\pathops{\suppath{\fpath}}}$ is
satisfiable. 
Similar to the case of sequential programs, one can check the
feasibility of $\fpath$ by checking the satisfiability of the path formula 
corresponding to the SSA form of $\pathops{\suppath{\fpath}}$.

\begin{exa} 
Suppose that the top path in Figure~\ref{fig:CexPath} is a
counter-example path (the target node of the last edge is an error
node).
The bottom path is the suppressed version of the top one.  
The dashed edge is an \ARF connector. 
To check feasibility of the path by means of satisfiability of the
corresponding path formula, we check the satisfiability of the
following formula:
\[
  \mathtt{x1 = x0 + y0} 
  \AND \mathtt{y1 = 7} 
  \AND \mathtt{x2 = z0} 
  \AND \mathtt{x2 < y1 + z0}.
\]

\begin{figure}
{\scriptsize
\psset{xunit=0.8cm,yunit=0.8cm,labelsep=3pt,nodesep=2pt,linewidth=2pt}
\begin{pspicture}(0,0)(17,3)

\pnode(1,2.5){N1}
\pnode(4,2.5){N2}
\pnode(7,2.5){N3}
\pnode(10,2.5){N4}
\pnode(13,2.5){N5}
\pnode(16,2.5){N6}

\pnode(2.5,0.5){M1}
\pnode(5.5,0.5){M2}
\pnode(8.5,0.5){M3}
\pnode(11.5,0.5){M4}
\pnode(14.5,0.5){M5}

\ncline{->}{N1}{N2}\nbput{\code{x := x+y}}
\ncline{->}{N2}{N3}\nbput{\code{y := 7}}
\ncline[linestyle=dashed]{->}{N3}{N4}
\ncline{->}{N4}{N5}\nbput{\code{x := z}}
\ncline{->}{N5}{N6}\nbput{\code{[x < y+z]}}

\ncline{->}{M1}{M2}\nbput{\code{x := x+y}}
\ncline{->}{M2}{M3}\nbput{\code{y := 7}}
\ncline{->}{M3}{M4}\nbput{\code{x := z}}
\ncline{->}{M4}{M5}\nbput{\code{[x < y+z]}}

\pnode(8.5,2){Top}
\pnode(8.5,1){Down}
\ncline[linewidth=5pt]{->}{Top}{Down}\naput{Suppressed}

\end{pspicture}}
\caption{An example of a counter-example path. \label{fig:CexPath}}
\end{figure}
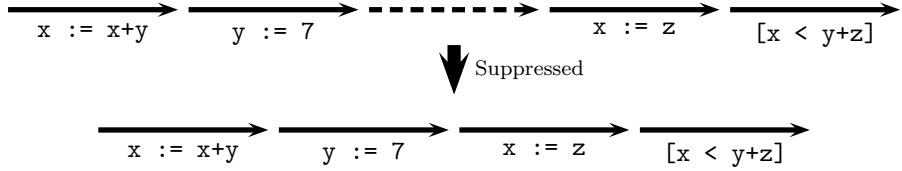
\end{exa}

\subsection{\ARF Refinement}
\label{subsec:ARFRefine}

When the counter-example path $\fpath$ is infeasible, we need to rule
out such a path by refining the precision of nodes in the \ARF.  
\ARF refinement amounts to finding additional predicates to refine the
precisions. 
Similar to the case of sequential programs, these additional
predicates can be extracted from the path formula corresponding 
to sequence $\pathops{\suppath{\fpath}}$ by using the Craig
interpolant refinement method described in Section~\ref{subsec:SWMC}.

As described in Section~\ref{subsec:ARF} newly discovered predicates 
can be added to precisions associated to locations, threads, or the global 
region.
Consider again the Craig interpolant method in Section~\ref{subsec:SWMC}.
Let $\tedge_1,\ldots,\tedge_n$ be the sequence of edges labelled by
the operations $op_1,\ldots,op_n$ of $\pathops{\suppath{\fpath}}$,
that is, for $i=1,\ldots,n$, the edge $\tedge_i$ is labelled by
$op_i$. Let $p$ be a predicate extracted from the interpolant $\psi^k$
of $(\varphi^{1,k},\varphi^{k+1,n})$ for $1 \leq k < n$, and let the nodes
\[ 
  (\ststate{l_1}{\varphi_1},
  \ldots,\ststate{l_i}{\varphi_i},\ldots,
  \ststate{l_N}{\varphi_N},
  \varphi,
  \sst)
\]
and
\[ 
  (\ststate{l_1}{\varphi'_1},
  \ldots,\ststate{l'_i}{\varphi'_i},\ldots,
  \ststate{l_N}{\varphi'_N},
  \varphi',
  \sst')
\] 
be, respectively, the source and target nodes of the edge $\tedge_k$
such that the running thread in the source node's scheduler state is
the thread $T_i$.
If $p$ contains only variables local to $T_i$, then we can add 
$p$ to the precision associated with the location $l'_i$, to the 
the precision associated with $T_i$, or to the precision associated with 
the global region.
Other precisions refinement strategies are applicable. For example,
one might add a predicate into the precision associated with the global region 
if and only if the predicate contains variables local to several threads.

Similar to the \ART refinement in the case of sequential programs,
once the precisions are refined, we refine the \ARF by removing the
infeasible counter-example path or by removing part of the \ARF that
contains the infeasible path, and then reconstruct again the \ARF
using the refined precisions.

\subsection{Havocked Operations}
\label{subsec:HavockedOp}

Computing the abstract strongest post-conditions with respect to the havocked 
operation in the rule~\ref{rule:ARFExp1} is necessary, not only to keep 
the regions of the \ARF node consistent, but, more importantly, to maintain 
soundness: never reports safe for an unsafe case.
Suppose that the region of a non-running thread $T$ is the formula $x=g$, 
where $x$ is a variable local to $T$ and $g$ is a shared global variable. 
Suppose further that the global region is $true$.
If the running thread $T'$ updates the value of $g$ with, for example,
the assignment $g := w$, for some variable $w$ local to $T'$, 
then the region $x=g$ of $T$ might no longer hold, and has to be invalidated. 
Otherwise, when $T$ resumes, and, for example, checks for an assertion 
\code{assert($x=g$)}, then no assertion violation can occur. 
One way to keep the region of $T$ consistent is to update the region
using the $\havoc(g := w)$ operation, as shown in the rule~\ref{rule:ARFExp1}.
That is, we compute the successor region of $T$ as 
$\ASP{x=g}{g := a}{\lprec{l}}$, where $a$ is a fresh variable and $l$
is the current location of $T$. The fresh variable $a$ essentially denotes
an arbitrary value that is assigned to $g$.

Note that, by using a $\havoc(op)$ operation, we do not leak variables
local to the running thread when we update the regions of non-running threads.
Unfortunately, the use of $\havoc(op)$ can cause loss of precision. One way 
to address this issue is to add predicates containing local and global 
variables to the precision associated with the global region. An alternative
approach, as described in~\cite{DBLP:conf/cav/DonaldsonKKW11}, is to simply 
use the operation $op$ (leaking the local variables) when updating 
the regions of non-running threads.

\subsection{Summary of ESST}
\label{subsec:SummaryESST}

\begin{figure}

\begin{tabular}{|c|c|} \hline

\begin{minipage}{0.48\textwidth}
\includegraphics{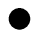} 
\end{minipage}
& 
\begin{minipage}{0.48\textwidth}
\includegraphics{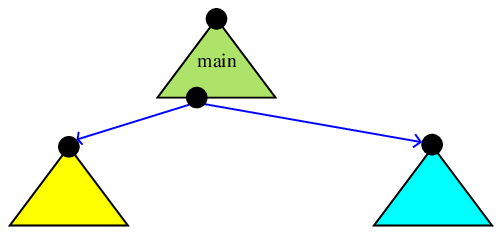} 
\end{minipage} \\ 

(a) & (b) \\ \hline

\begin{minipage}{0.48\textwidth}
\includegraphics{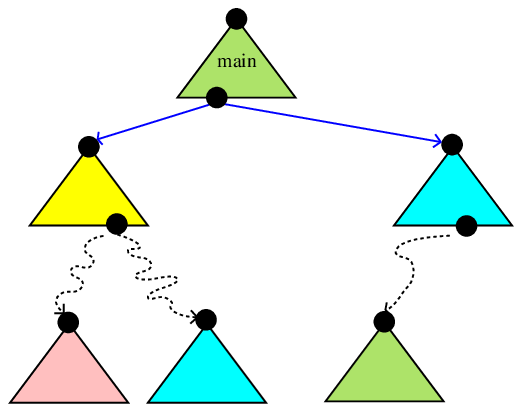} 
\end{minipage}
& 
\begin{minipage}{0.48\textwidth}
\includegraphics{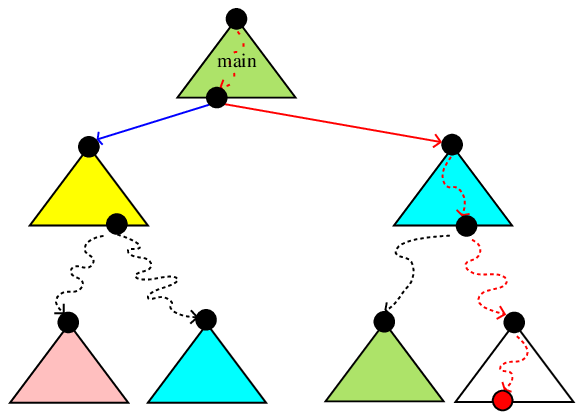} 
\end{minipage}  \\

(c) & (d) \\ \hline

\begin{minipage}{0.48\textwidth}
\includegraphics{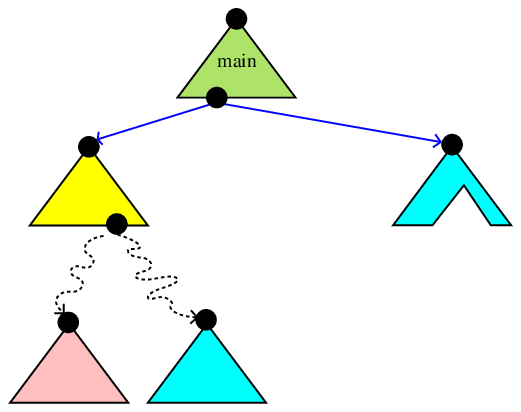} 
\end{minipage}
& 
\begin{minipage}{0.48\textwidth}
\includegraphics{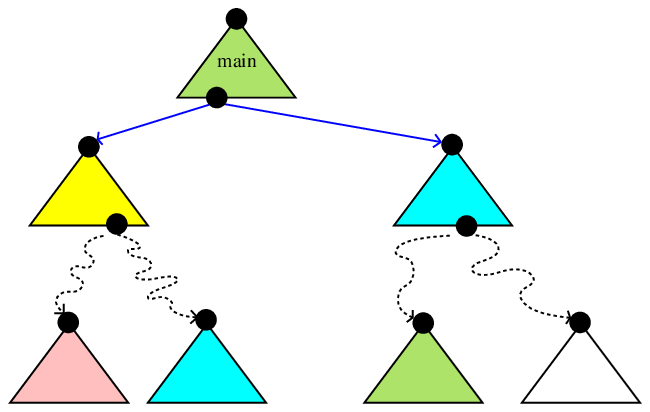} 
\end{minipage} \\

(e) & (f) \\ \hline

\end{tabular}

\caption{\ARF construction in \ESST. \label{fig:ESST}}
\end{figure}

The \ESST algorithm takes a threaded program $P$ as an input
and, when its execution terminates, returns either a feasible
counter-example path and reports that $P$ is unsafe, or a safe 
\ARF and reports that $P$ is safe. The execution of
$\ESST(P)$ can be illustrated in Figure~\ref{fig:ESST}:
\begin{enumerate}[(1)]
  \item Start with an \ARF consisting only of the initial node,
  as shown in Figure~\ref{fig:ESST}(a). 

  \item Pick an \ARF node that can be expanded and apply 
  the rules~\ref{rule:ARFExp1} or ~\ref{rule:ARFExp2} to grow the \ARF, 
  as shown in Figures~\ref{fig:ESST}(b) and ~\ref{fig:ESST}(c).
  The different colors denote the different threads to which 
  the \ARTs belong.

  \item If we reach an error node, as shown by the red line 
  in Figure~\ref{fig:ESST}(d), we analyze the counter-example path.
  
  \begin{enumerate}[(a)]
    \item If the path is feasible, then report that $P$ is 
    \emph{unsafe}.
    \item If the path is spurious, then refine the \ARF:
    \begin{enumerate}[(i)]
      \item Discover new predicates to refine abstractions.
      \item Undo part of the \ARF, as shown in Figure~\ref{fig:ESST}(e).
      \item Goto (2) to reconstruct the \ARF.
    \end{enumerate}
  \end{enumerate}

  \item If the \ARF is safe, as shown in Figure~\ref{fig:ESST}(f), then
  report that $P$ is \emph{safe}.
\end{enumerate}

\subsection{Correctness of ESST}
\label{subsec:CorrectESST}

To prove the correctness of \ESST, we need to introduce several notions
and notations that relate the \ESST algorithm with the operational 
semantics in Section~\ref{sec:ProgModel}. Given two states $\st_1$
and $\st_2$ whose domains are disjoint, we denote by $\st_1 \cup \st_2$
the union of two states such that 
$\StDom{\st_1 \cup \st_2}$ is $\StDom{\st_1} \cup \StDom{\st_2}$, and, 
for every $x \in \StDom{\st_1 \cup \st_2}$, we have
\[
  (\st_1 \cup \st_2)(x) 
      = \left\{ \begin{array}{ll}
                  \st_1(x) & \mbox{if $x \in \StDom{\st_1}$;} \\ 
                  \st_2(x) & \mbox{otherwise.}
                \end{array} \right.
\]
Let $P$ be a threaded program with $N$ threads, and $\cg$ be 
a configuration 
\[
    \conf{(l_1,\st_1),\ldots,(l_N,\st_N)}{\gst}{\sst},
\]
of $P$. Let $\frnode$ be an \ARF node 
\[
  (\ststate{l'_1}{\varphi_1},\ldots,\ststate{l'_N}{\varphi_N},\varphi,\sst'), 
\]
for $P$. We say that the configuration $\cg$ \emph{satisfies} the \ARF node
$\frnode$, denoted by $\cg \models \frnode$ if and only if 
for all $i=1,\ldots,N$, we have $l_i = l'_i$ and $\st_i \cup \gst
\models \varphi_i$,
$\bigcup_{i=1,\ldots,N} \st_i \cup \gst \models \varphi$, and 
$\sst = \sst'$.

By the above definition, it is easy to see that, for any initial
configuration $\cg_0$ of $P$, we have $\cg_0 \models \frnode_0$ for
the initial \ARF node $\frnode_0$.
In the sequel we refer to the configurations of $P$ and 
the \ARF nodes (or connectors) for $P$ when we speak about configurations 
and \ARF nodes (or connectors), respectively.

We now show that the node expansion rules~\ref{rule:ARFExp1} and
\ref{rule:ARFExp2} create successor nodes that are over-approximations
of the configurations reachable by performing operations considered in
the rules.

\begin{lem}\label{lem:Overapprox} 
Let $\frnode$ and $\frnode'$ be \ARF nodes for a threaded program $P$
such that $\frnode'$ is a successor node of $\frnode$.  Let $\cg$ be a
configuration of $P$ such that $\cg \models \frnode$.  The following
properties hold:
\begin{enumerate}[\em(1)]
  \item If $\frnode'$ is obtained from $\frnode$ by 
  the rule~\ref{rule:ARFExp1} with the performed operation $op$, then,
  for any configuration $\cg'$ of $P$ such that $\cg \transsem{op} \cg'$,
  we have $\cg' \models \frnode'$.
  
  \item If $\frnode'$ is obtained from $\frnode$ by 
  the rule~\ref{rule:ARFExp2}, then, for any configuration $\cg'$ of $P$ 
  such that $\cg \transsem{\cdot} \cg'$ and the scheduler states 
  of $\frnode'$ and $\cg'$ coincide, we have $\cg' \models \frnode'$.
\end{enumerate}
\end{lem}
\newcommand{\ProofLemOverapprox}{
\proof
We first prove property (1). Let $\frnode$ and $\frnode'$ be as 
follows:
\[
\begin{array}{lcl}
 \frnode & = & (\ststate{l_1}{\varphi_1},
                \ldots,
                \ststate{l_i}{\varphi_i},
                \ldots
                \ststate{l_N}{\varphi_N},
                \varphi,
                \sst) \\
 \frnode'& = & (\ststate{l_1}{\varphi'_1},
                \ldots,
                \ststate{l'_i}{\varphi'_i},
                \ldots
                \ststate{l_N}{\varphi'_N},
                \varphi',
                \sst'), 
\end{array}
\]
such that 
$\sst(\tst{T_i}) = \RUNNING$ and for all $j\neq i$, we have
$\sst(\tst{T_j}) \neq \RUNNING$.  
Let $\TCFG{T_i}=(L,E,l_0,L_{err})$ be the \CFG for $T_i$ such that 
$(l_i,op,l'_i) \in E$.
Let $\cg$ and $\cg'$ be as follows: 
\[ 
\begin{array}{lcl}
  \cg  & = & \conf{(l_1,\st_1),
                   \ldots,
                   (l_i,\st_i),
                   \ldots,
                   (l_N,\st_N)}{\gst}{\sst} \\

  \cg' & = & \conf{(l_1,\st_1),
                   \ldots,
                   (l'_i,\st'_i),
                   \ldots,
                   (l_N,\st_N)}{\gst'}{\sst''},
\end{array}
\]
such that $\cg \transsem{op} \cg'$. We need to prove that 
$\cg' \models \frnode'$. 
Let $\hat{op}$ be $op$ if $op$ contains no primitive function call, or
be $op'$ as in the second case of the rule~\ref{rule:ARFExp1}. 
First, from $\cg \models \frnode$, we have $\st_i \cup \gst \models
\varphi_i$. 
By the definition of operational semantics of $\hat{op}$ and the
definition of $\SP{\varphi_i}{\hat{op}}$, it follows that $\st'_i \cup
\gst' \models \SP{\varphi_i}{\hat{op}}$. 
Since $\SP{\varphi_i}{\hat{op}}$ implies
$\ASP{\varphi_i}{\hat{op}}{\pi}$ for any precision $\pi$, and
$\varphi'_i$ is $\ASP{\varphi_i}{\hat{op}}{\lprec{l'_i}}$ for some
precision $\lprec{l'_i}$ associated with $l'_i$, it follows that
$\st'_i \cup \gst' \models \varphi'_i$. 
A similar reasoning can be applied to prove that $\st'_j \cup \gst'
\models \varphi'_j$ for $j \neq i$ and $ \bigcup_{i=1,\ldots,N} \st'_i 
\cup \gst' \models \varphi'$.
We remark that the $\havoc(\hat{op})$ operation only makes the values
of global variables possibly assigned in $\hat{op}$ unconstrained.
To prove that $\cg' \models \frnode'$, it remains to show that $\sst'$
and $\sst''$ coincide. 
Now, consider the case where $\hat{op}$ does not contain any call to
primitive function.  
It is then trivial that $\sst' = \sst''$. 
Otherwise, if $\hat{op}$ contains a call to primitive function, then,
since the primitive executor follows the operational semantics, that
is, $\Sexec(\sst,f(\vec{x}))$ computes
$\sem{f(\vec{x})}(\cdot,\cdot,\sst)$, we have $\sst' = \sst''$. 
Hence, we have proven that $\cg' \models \frnode'$.

For property (2), let $\frnode$ and $\frnode'$ be as 
follows:
\[
\begin{array}{lcl}
 \frnode & = & (\ststate{l_1}{\varphi_1},
                \ldots,
                \ststate{l_N}{\varphi_N},
                \varphi,
                \sst) \\
 \frnode'& = & (\ststate{l_1}{\varphi_1},
                \ldots,
                \ststate{l_N}{\varphi_N},
                \varphi,
                \sst'), 
\end{array}
\]
such that $\sst(\tst{T_i}) \neq \RUNNING$ for all $i = 1,\ldots,N$. 
Let $\cg$ and $\cg'$ be as follows: 
\[ 
\begin{array}{lcl}
  \cg  & = & \conf{(l_1,\st_1),
                   \ldots,
                   (l_N,\st_N)}{\gst}{\sst} \\

  \cg' & = & \conf{(l_1,\st'_1),
                   \ldots,
                   (l_N,\st'_N)}{\gst'}{\sst''},
\end{array}
\]
By the operational semantics, we have $\st_i = \st'_i$ for all $i=1,\ldots,N$,
and $\gst = \gst'$. Since $\sst' = \sst''$, it follows from 
$\cg \models \frnode$ that $\cg' \models \frnode'$.  
\qed 
}

Let $\tedge$ be an \ART edge with source node 
\[
 \frnode = (\ststate{l_1}{\varphi_1},
            \ldots,
            \ststate{l_i}{\varphi_i},
            \ldots
            \ststate{l_N}{\varphi_N},
            \varphi,
            \sst) 
\]
and target node 
\[
 \frnode' = (\ststate{l_1}{\varphi'_1},
             \ldots,
             \ststate{l'_i}{\varphi'_i},
             \ldots
             \ststate{l_N}{\varphi'_N},
             \varphi',
             \sst'), 
\]
such that 
$\sst(\tst{T_i}) = \RUNNING$ and for all $j\neq i$, we have
$\sst(\tst{T_j}) \neq \RUNNING$.
Let $\TCFG{T_i}=(L,E,l_0,L_{err})$ be the \CFG for $T_i$ such that 
$(l_i,op,l'_i) \in E$.
Let $\cg$ and $\cg'$ be configurations. We denote by 
$\cg \simtrans{\tedge} \cg'$ if $\cg \models \frnode$, $\cg' \models \frnode'$,
and $\cg \transsem{op} \cg'$. 
Note that, the operation $op$ is the operation labelling the edge of
\CFG, not the one labelling the \ART edge $\tedge$. 
Similarly, we denote by $\cg \simtrans{\fconn} \cg'$ for an \ARF
connector $\fconn$ if $\cg \models \frnode$, $\cg' \models \frnode'$,
and $\cg \transsem{\cdot} \cg'$. Let $\fpath = \xi_1,\ldots,\xi_m$ be
an \ARF path. 
That is, for each $i = 1,\ldots,m$, the element $\xi_i$ is either an
\ART edge or an \ARF connector. 
We denote by $\cg \simtrans{\fpath} \cg'$ if there exists a
computation sequence $\cg_1,\ldots,\cg_{m+1}$ such that $\cg_i
\simtrans{\xi_i} \cg_{i+1}$ for all $i=1,\ldots,m$, and $\cg = \cg_1$
and $\cg' = \cg_{m+1}$.

In Section~\ref{sec:ProgModel} the notion of strongest post-condition
is defined as a set of reachable states after executing some
operation.  We now try to relate the notion of configuration with the
notion of strongest post-condition. Let $\cg$ be a configuration
\[
    \cg = \conf{(l_1,\st_1),\ldots,(l_i,\st_i),\ldots,(l_N,\st_N)}{\gst}{\sst},
\]
and $\varphi$ be a formula whose free variables range over
$\bigcup_{k=1,\ldots,N} \StDom{\st_k} \cup \StDom{\gst}$.
We say that the configuration \emph{satisfies} the formula $\varphi$,
denoted by $\cg \models \varphi$ if $\bigcup_{k=1,\ldots,N} 
\st_k \cup \gst \models \varphi$.
Suppose that in the above configuration $\cg$ we have $\sst(\tst{T_i})
= \RUNNING$ and $\sst(\tst{T_j}) \neq \RUNNING$ for all $j\neq i$.
Let $\TCFG{T_i}=(L,E,l_0,L_{err})$ be the \CFG for $T_i$ such that
$(l_i,op,l'_i) \in E$. Let $\hat{op}$ be $op$ if $op$ does not contain
any primitive function call, otherwise $\hat{op}$ be $op'$ as in the
second case of the expansion rule~\ref{rule:ARFExp1}. Then, for any
configuration
\[
  \cg' = \conf{(l_1,\st_1),\ldots,(l'_i,\st'_i),\ldots,(l_N,\st_N)}{\gst'}{\sst'},
\]
such that $\cg \transsem{op} \cg'$, we have $\cg' \models
\SP{\varphi}{\hat{op}}$.  
Note that, the scheduler states $\sst$ and $\sst'$ are not constrained
by, respectively, $\varphi$ and $\SP{\varphi}{\hat{op}}$, and so they can
be different.

When $\ESST(P)$ terminates and reports that $P$ is safe, we require
that, for every configuration $\cg$ reachable in $P$, there is a node
in $\forest$ such that the configuration satisfies the node.  
We denote by $\ReachC{P}$ the set of configurations reachable in $P$,
and by $\NodesF{\forest}$ the set of nodes in $\forest$.

\begin{thm}[Correctness]\label{thm:CorrectESST}
Let $P$ be a threaded program. For every terminating execution of
$\ESST(P)$, we have the following properties:
\begin{enumerate}[\em(1)]
  \item If $\ESST(P)$ returns a feasible counter-example path $\fpath$, 
        then we have $\cg \simtrans{\fpath} \cg'$ for an initial
        configuration $\cg$ and an error configuration $\cg'$ of $P$.
  \item If $\ESST(P)$ returns a safe \ARF $\forest$, then for every 
        configuration $\cg \in \ReachC{P}$, there is an \ARF node
        $\frnode \in \NodesF{\forest}$ such that $\cg \models
        \frnode$.
\end{enumerate}
\end{thm}
\newcommand{\ProofThmCorrectESST}{
\proof
We first prove property (1). Let the counter-example path $\fpath$ be
the sequence $\xi_1,\ldots,\xi_m$, such that, for each $i =
1,\ldots,m$, the element $\xi_i$ is either an \ART edge or an \ARF
connector.  
We need to show the existence of a computation sequence
$\cg_1,\ldots,\cg_{m+1}$ such that $\cg_i \simtrans{\xi_i} \cg_{i+1}$
for all $i=1,\ldots,m$, and $\cg = \cg_1$ and $\cg' = \cg_{m+1}$.
Let $\fpath^j$, for $0 \leq j \leq m$, denote the prefix
$\xi_1,\ldots,\xi_j$ of $\fpath$. 
Let $\psi^j$ be the strongest post-condition after performing the
operations in the suppressed version of $\fpath^j$. 
That is, $\psi^j$ is $\SP{true}{\pathops{\suppath{\fpath^j}}}$. 
For $k=1,\ldots,m$, we need to show that, for any configuration
$\cg_k$ satisfying $\psi^{k-1}$ and the source node of $\xi_k$, there
is a configuration $\cg_{k+1}$ such that $\cg_{k+1}$ satisfies
$\psi^k$ and the target node of $\xi_k$.

First, any configuration satisfies $true$, and thus $\cg \models true$.  
By definition of counter-example path, the source node of $\xi_1$ is
an initial node $\frnode_0$. 
Any initial configuration satisfies the initial node, and thus $\cg
\models \frnode_0$.  
Second, take any $1 \leq k \leq m$, and assume that we have a
configuration $\cg_k$ satisfying $\psi^{k-1}$ and the source node of
$\xi_k$.
Consider the case where $\xi_k$ is an \ART edge obtained by unwinding
\CFG edge labelled by an operation $op$. 
Let $\hat{op}$ be the label of the \ART edge. 
That is, $\hat{op} = op$ if $op$ has no primitive function call;
otherwise $\hat{op} = op'$ where $op'$ is defined in the second case
of rule~\ref{rule:ARFExp1}. 
Since $\psi^m$ is satisfiable, then so is $\psi^k$. 
It means that there is a configuration $\cg'$ such that $\cg_k
\transsem{\hat{op}} \cg'$ and $\cg' \models \psi^k$.  
Recall that the scheduler state of $\cg_{k+1}$ is not constrained by
$\psi^k$ and primitive function calls can only modify scheduler
states. 
Thus, there is a configuration $\cg_{k+1}$ that differs from $\cg'$
only in the scheduler state, such that $\cg_k \transsem{op} \cg_{k+1}$
and $\cg_{k+1} \models \psi^k$.
When $op$ has no primitive function call, then we simply take $\cg'$
as $\cg_{k+1}$.  
By Lemma~\ref{lem:Overapprox}, it follows that $\cg_{k+1}$ satisfies
the target node of $\xi_k$, and hence we have $\cg_k \simtrans{\xi_k}
\cg_{k+1}$, as required.

Consider now the case where $\xi_k$ is an \ARF connector. 
The connector $\xi_k$ is suppressed in the computation of the
strongest post-condition, that is $\psi^{k}$ is $\psi^{k-1}$. 
We obtain $\cg_{k+1}$ from $\cg_k$ by replacing $\cg_k$'s scheduler
state with the scheduler state in the target node of $\xi_k$. 
Since free variables of $\psi^k$ do not range over variables tracked by
the scheduler state and $\cg_k \models \psi^{k-1}$, we have $\cg_{k+1}
\models \psi^k$. 
By the construction of $\cg_{k+1}$ and by Lemma~\ref{lem:Overapprox},
it follows that $\cg_{k+1}$ satisfies the target node of $\xi_k$, and
hence we have $\cg_k \simtrans{\xi_k} \cg_{k+1}$, as required.

We now prove property (2). We prove that, for any run
$\cg_0,\cg_1,\ldots$ of $P$ and for any configuration $\cg_i$ in the
run, there is a node $\frnode \in \NodesF{\forest}$ such that $\cg_i
\models \frnode$. We prove the property by induction on the length $l$
of the run:
\begin{desCription}
\item\noindent{\hskip-12 pt\bf Case $l=1$:}\ This case is trivial
  because the initial configuration $\cg_0$ satisfies the initial
  node, and the construction of an \ARF starts with the initial node.
\item\noindent{\hskip-12 pt\bf Case $l>1$:}\ Let $\frnode \in
  \NodesF{\forest}$ be an \ARF node such that the configuration $\cg_n
  \models \frnode$.  If $\frnode$ is covered by another node $\frnode'
  \in \NodesF{\forest}$, then, by Definition~\ref{def:NodeCoverage} of
  node coverage, we have $\cg_l \models \frnode'$.
      Thus, we pick such an \ARF node $\frnode$ such that it is not
      covered by other nodes.

      Consider the transition $\cg_l \transsem{op} \cg_{l+1}$ from 
      $\cg_l$ to $\cg_{l+1}$. By the rule~\ref{rule:ARFExp1}, the node $\frnode$
      has a successor node $\frnode'$ obtained by performing the operation 
      $op$. By Lemma~\ref{lem:Overapprox}, we have $\cg_{l+1} \models \frnode'$,
      as required.

      Now, consider the transition $\cg_l \transsem{\cdot}
      \cg_{l+1}$. Because the scheduler \Sched implements the function
      $\SchedFn$ in the operational semantics, then, by the
      rule~\ref{rule:ARFExp2}, the node $\frnode$ has a successor node
      $\frnode'$ whose scheduler state coincide with $\cg_{l+1}$. By
      Lemma~\ref{lem:Overapprox}, we have
      $\cg_{l+1} \models \frnode'$, as required. \qed
\end{desCription}
}

%% file: po-for-esst.tex
\section{ESST + Partial-Order Reduction}
\label{sec:ESSTPlusPOR}

The \ESST algorithm often has to explore a large number of possible 
thread interleavings. However, some of them might be redundant because 
the order of interleavings of some threads is irrelevant. 
Given $N$ threads such that each of them accesses a disjoint set of 
variables, there are $N!$ possible interleavings that \ESST has to explore. 
The constructed \ARF will consists of $2^N$ abstract states (or nodes). 
Unfortunately, the more abstract states to explore, the more computations 
of abstract strongest post-conditions are needed, and the more coverage
checks are involved.
Moreover, the more interleavings to explore, the more possible spurious 
counter-example paths to rule out, and thus the more refinements are needed.
As refinements result in keeping track of additional predicates, 
the computations of abstract strongest post-conditions become expensive.
Consequently, exploring all possible interleavings degrades the performance 
of \ESST and leads to state explosion.

Partial-order reduction techniques
(\POR)~\cite{DBLP:books/sp/Godefroid96,PeledCAV93,ValmariAPN90} have
been successfully applied in explicit-state software model checkers
like \spin~\cite{DBLP:journals/ac/Holzmann05} and
\verisoft~\cite{DBLP:journals/fmsd/Godefroid05} to avoid exploring
redundant interleavings.
\POR has also been applied to symbolic model checking techniques as
shown
in~\cite{DBLP:conf/cav/KahlonGS06,DBLP:conf/tacas/WangYKG08,DBLP:journals/fmsd/AlurBHQR01}.
In this section we will extend the \ESST algorithm with \POR techniques.
However, as we will see, such an integration is not trivial 
because we need to ensure that in the construction of the \ARF 
the \POR techniques do not make \ESST unsound.

\subsection{Partial-Order Reduction (POR)}
\label{subsec:POR}

Partial-order reduction (\POR) is a
model checking technique that is aimed at combating the state
explosion by exploring only representative subset of all possible
interleavings. 
\POR exploits the commutativity of concurrent transitions that result
in the same state when they are executed in different orders. 

We present \POR using the standard notions and notations used 
in~\cite{DBLP:books/sp/Godefroid96,ClarkeGrumbergPeledBook99}. 
We model a concurrent program as a transition system
$M=(S,S_0,T)$, where $S$ is the finite set of states, $S_0 \subset S$
is the set of initial states, and $T$ is a set of transitions such
that for each $\transa \in T$, we have $\transa \subset S \times
S$. 
We say that $\transaSS{s}{s'}$ holds and often write it as
$\StransaS{s}{s'}$ if $(s,s') \in \transa$. 
A state $s'$ is a successor of a state $s$ if $\StransaS{s}{s'}$ for
some transition $\transa \in T$. In the following we will only consider 
deterministic transitions, and often write $s' = \transaS{s}$ 
for $\transaSS{s}{s'}$.
A transition $\transa$ is \emph{enabled} in a state $s$ 
if there is a state $s'$ such that $\transaSS{s}{s'}$ holds. The set of 
transitions enabled in a state $s$ is denoted by $\enabled{s}$.
A \emph{path} from a state $s$ in a transition system is a finite or
infinite sequence $s_0 \transX{\transa_0} s_1 \transX{\transa_1}
\cdots$ such that $s = s_0$ and $s_i \transX{\transa_i} s_{i+1}$ for
all $i$.
A path is empty if the sequence consists only of a single state.
The length of a finite path is the number of transitions in the path.

Let $M=(S,S_0,T)$ be a transition system, we denote by $\Reach{S_0}{T}
\subseteq S$ the set of states reachable from the states in $S_0$ by
the transitions in $T$: for a state $s \in \Reach{S_0}{T}$, there is a
finite path $s_0 \transX{\transa_0} \ldots \transX{\transa_{n-1}} s_n$
system such that $s_0 \in S_0$ and $s = s_n$.
In this work we are interested in verifying safety properties in the form 
of program assertion. To this end, we assume that there is a set 
$\Terr \subseteq T$ of \emph{error transitions} such that the set
\[
   \Err{M}{\Terr} = \setof{s \in S \mid \exists s'\in S. \exists \transa \in \Terr.\ \ \transaSS{s'}{s} \mbox{ holds } }
\]
is the set of \emph{error states} of $M$ with respect to $\Terr$.
A transition system $M = (S,S_0,T)$ is \emph{safe with respect to the
set $\Terr \subseteq T$ of error transitions} iff $\Reach{S_0}{T} \cap
\Err{M}{\Terr} = \emptyset$.

Selective search in \POR exploits the commutativity of concurrent
transitions. The concept of commutativity of concurrent transitions 
can be formulated by defining an independence relation on pairs 
of transitions. 

\begin{defi}[Independence Relation, Independent Transitions]
An \emph{independence relation} $I \subseteq T \times T$ is a
symmetric, anti-reflexive relation such that for each state $s \in S$
and for each $(\transa,\transb) \in I$ the following conditions are
satisfied:
\begin{desCription}
  \item\noindent{\hskip-12 pt\bf Enabledness:}\ If $\transa$ is in $\enabled{s}$, then $\transb$ is in
  $\enabled{s}$ iff $\transb$ is in $\enabled{\transaS{s}}$. 
  \item\noindent{\hskip-12 pt\bf Commutativity:}\ If $\transa$ and $\transb$ are in $\enabled{s}$, then 
  $\transaS{\transbS{s}} = \transbS{\transaS{s}}$.
\end{desCription}
We say that two transitions $\transa$ and $\transb$ are
\emph{independent} of each other if for every state $s$ they satisfy
the enabledness and commutativity conditions. 
We also say that two transitions $\transa$ and $\transb$ are
\emph{independent in a state $s$} of each other if they satisfy the
enabledness and commutativity conditions in $s$. 
\end{defi}

In the sequel we will use the notion of valid dependence relation to
select a representative subset of transitions that need to be
explored.

\begin{defi}[Valid Dependence Relation]
A \emph{valid dependence relation} $D \subseteq T \times T$ is a
symmetric, reflexive relation such that for every $(\transa,\transb) \not\in
D$, the transitions $\transa$ and $\transb$ are independent of each
other. 
\end{defi}

\subsubsection{The Persistent Set Approach}

To reduce the number of possible interleavings, in every state visited
during the state space exploration one only explores a representative
subset of transitions that are enabled in that state. 
However, to select such a subset we have to avoid possible
dependencies that can happen in the future. To this end, we appeal to
the notion of persistent set~\cite{DBLP:books/sp/Godefroid96}.

\begin{defi}[Persistent Set] 
A set $P \subseteq T$ of enabled transitions in a state $s$ is
\emph{persistent} in $s$ if for every finite non-empty path
$
  s = s_0 \transX{\transa_0} s_1 \transX{\transa_1} 
  \cdots 
  \transX{\transa_{n-1}} s_n \transX{\transa_n} s_{n+1}
$
such that $\transa_i \not\in P$ for all $i=0,\ldots,n$, we have 
$\transa_n$ independent of any transition in $P$ in $s_n$. 
\end{defi}

Note that the persistent set in a state is not unique. 
To guarantee the existence of successor state, we impose the
\emph{successor-state} condition on the persistent set: the persistent
set in $s$ is empty iff so is $\enabled{s}$. 
In the sequel we assume persistent sets satisfy the successor-state
condition.
We say that a state $s$ is \emph{fully expanded} if the persistent set
in $s$ equals $\enabled{s}$.
It is easy to see that, for any transition $\transa$ not in the
persistent set $P$ in a state $s$, the transition $\transa$ is
disabled in $s$ or independent of any transition in $P$.

We denote by $\ReachRed{S_0}{T} \subseteq S$ the set of states
reachable from the states in $S_0$ by the transitions in $T$ such that,
during the state space exploration, in every visited state we only
explore the transitions in the persistent set in that state.
That is, for a state $s \in \ReachRed{S_0}{T}$, there is a finite path
$s_0 \transX{\transa_0} \ldots \transX{\transa_{n-1}} s_n$
in the transition system such that $s_0 \in S_0$ and $s = s_n$,
and $\transa_i$ is in the persistent set of $s_i$, for $i = 0,\ldots,n-1$.
It is easy to see that $\ReachRed{S_0}{T} \subseteq \Reach{S_0}{T}$. 

To preserve safety properties of a transition system, we need to
guarantee that the reduction by means of persistent sets does not remove 
all interleavings that lead to an error state.
To this end, we impose the \emph{cycle condition} on
$\ReachRed{S_0}{T}$~\cite{ClarkeGrumbergPeledBook99,PeledCAV93}: a
cycle is not allowed if it contains a state in which a transition
$\transa$ is enabled, but $\transa$ is never included in the
persistent set of any state $s$ on the cycle.
That is, if there is a cycle 
$s_0 \transX{\transa_0} \ldots \transX{\transa_{n-1}} s_n = s_0$ induced
by the states $s_0,\ldots,s_{n-1}$ in $\ReachRed{S_0}{T}$ such that
$\transa_i$ is persistent in $s_i$, for $i = 0,\ldots,n-1$ and 
$\transa \in \enabled{s_j}$ for some $0 \leq j < n$, then $\transa$
must be in the persistent set of any of $s_0,\ldots,s_{n-1}$.

\begin{thm}\label{thm:PORCorrect} 
A transition system $M=(S,S_0,T)$ is safe w.r.t. a set $\Terr
\subseteq T$ of error transitions iff $\ReachRed{S_0}{T}$ that
satisfies the cycle condition does not contain any error state from
$\Err{M}{\Terr}$.
\end{thm} 
\newcommand{\ProofThmPORCorrect}{
\proof

If the transition system $M$ is safe w.r.t. $\Terr$, then
$\ReachRed{S_0}{T} \cap \Err{M}{\Terr} = \emptyset$ follows obviously
because $\Reach{S_0}{T} \cap \Err{M}{\Terr} = \emptyset$ and
$\ReachRed{S_0}{T} \subseteq \Reach{S_0}{T}$.

For the other direction, let us assume the transition system $M$ being
unsafe w.r.t. $\Terr$. Without loss of generality we also assume that
$\Terr = \setof{\transa}$.
We prove that for every state $s_0 \in S$ such that there is a path of
length $n > 0$ leading to an error state $s_e$, then there is a path
from $s_0$ to an error state $s'_e$ such that the path consists only
of transitions in the persistent sets of visited states. 
When the state $s_0$ is in $S_0$, then the states visited by the
latter path are only states in $\ReachRed{S_0}{T}$. 
We first show the proof for $n=1$ and $n=2$, and then we generalize it
for arbitrary $n>1$.

\begin{desCription}
  \item\noindent{\hskip-12 pt\bf Case $n=1$:}\ Let $s_0 \in S$ be such that $\StransaS{s_0}{s_e}$ holds
  for an error state $s_e$. By the successor-state condition, the persistent
  set in $s_0$ is non-empty. If the only persistent set in $s_0$ is 
  the singleton set $\setof{\transa}$, then the path $\StransaS{s_0}{s_e}$
  is the path leading to an error state and the path consists only of 
  transitions in the persistent sets of visited states. 
  Suppose that the transition $\transa$ is not in the persistent set in 
  $s_0$. Take the greatest $m > 0$ such that there is a path
  \[
      s_0 \transX{\transc_0} s_1  \transX{\transc_1} 
     \cdots 
     \transX{\transc_{m-1}} s_m,
  \] 
  where for all $i=0,\ldots,m-1$, the set $P_i$ is the persistent set
  in state $s_i$, the transition $\transc_i$ is in $P_i$, and the
  transition $\transa$ in not in $P_i$ (see Figure~\ref{fig:SafetyPreserve1}).
  First, the above path exists because of the successor-state condition 
  and  it must be finite because the set $S$ of states is finite. The path 
  cannot form a cycle, otherwise by the cycle condition the transition 
  $\transa$ will have been in the persistent set in one of the states that 
  form the cycle. That is, by the above path, we delay the exploration of 
  $\transa$ as long as possible.
  Second, since the transition $\transa$ is enabled in $s_0$ and 
  is independent in $s_i$ of any transition in $P_i$ for all 
  $i=0,\ldots,m-1$ (otherwise $P_i$ is not a persistent set), then 
  $\transa$ remains enabled in $s_j$ for $j = 1,\ldots,m$.
  Third, since $m$ is the greatest number, we have $\transa$ in the
  persistent set in the state $s_m$, and furthermore
  $\StransaS{s_m}{s'_e}$ holds for an error state $s'_e$.
  Thus, the path 
  \[
      s_0 \transX{\transc_0} 
      \cdots 
      \transX{\transc_{m-1}} s_m \transX{\transa} s'_e 
  \] 
  is the path from $s_0$ leading to an error state $s'_e$ involving only
  transitions in the persistent sets of visited states.

  \item\noindent{\hskip-12 pt\bf Case $n=2$:}\ Let $s_0 \in S$ be such that there is a path
  \[
     s_0 \transX{\transb_0} s'_1 \transX{\transb_1 = \transa} s_e
  \] 
  for some state $s'_1$ and an error state $s_e$. By the
  successor-state condition, the persistent set in $s_0$ is
  non-empty. If the only persistent set in $s_0$ is the singleton set
  $\setof{\transb_0}$, then the path $s_0 \transX{\transb_0} s'_1$
  consists only of transition in the persistent set. By the case
  $n=1$, it is guaranteed that there is a path from $s'_1$ leading to
  an error state $s'_e$ such that the path consists only of
  transitions in the persistent sets of visited states. Thus, there is
  a path from $s_0$ leading to an error state $s'_e$ such that the
  path consists only of transitions in the persistent sets of visited
  states.

  Suppose that the transition $\transb_0$ is not in the persistent set in 
  $s_0$. Take the greatest $m > 0$ such that there is a path
  \[
      s_0 \transX{\transc_0} s_1  \transX{\transc_1} 
     \cdots 
     \transX{\transc_{m-1}} s_m,
  \] 
  where for all $i=0,\ldots,m-1$, the set $P_i$ is the persistent set
  in state $s_i$, the transition $\transc_i$ is in $P_i$, and the
  transition $\transb_0$ in not in $P_i$ (see 
  Figure~\ref{fig:SafetyPreserve2}). With the same reasoning as in
  the case of $n=1$, the above path exists, and is finite and acyclic. 
  That is, we delay the exploration of $\transb_0$ as long as possible. 

  Consider now the path 
  \[
      s_0 \transX{\transc_0} s_1  \transX{\transc_1} 
     \cdots 
     \transX{\transc_{m-1}} s_m \transX{\transb_0} s'_{m+1}.
  \] 
  We show that an error state is reachable from the state $s'_{m+1}$.
  First, since the transitions $\transc_0$ and $\transb_0$ are independent
  in $s_0$, the transitions $\transc_0$ and $\transb_0$ are enabled,
  respectively, in the states $s'_1$ and $s_1$, and they commute in the  
  state $s'_2$. The transition $\transc_0$ is also independent of 
  the transition $\transa$ in $s'_1$, otherwise $P_0$ is not a persistent
  set in $s_0$. Thus, the transition $\transa$ is enabled in $s'_2$. 
  Second, since the transitions $\transc_1$ and $\transb_0$ are independent
  in $s_1$, the transitions $\transc_1$ and $\transb_0$ are enabled,
  respectively, in the states $s'_2$ and $s_2$, and they commute in the  
  state $s'_3$. The transition $\transc_1$ is independent of the transition 
  $\transa$ in $s'_2$, otherwise $P_1$ is not a persistent set in $s_1$. 
  Thus, the transition $\transa$ is enabled in $s'_3$.

  By repeatedly applying the above reasoning, it follows that the transition
  $\transa$ is enabled in the state $s'_{m+1}$. If the singleton set 
  $\setof{\transa}$ is the only persistent set in $s'_{m+1}$, then we are 
  done. That is, the path
  \[
      s_0 \transX{\transc_0} s_1  \transX{\transc_1} 
     \cdots 
     \transX{\transc_{m-1}} s_m \transX{\transb_0} s'_{m+1} 
     \transX{\transa} s'_e
  \] 
  is the path from $s_0$ leading to an error state $s'_e$ such that it
  consists only of transitions in the persistent sets of visited states.

  In the same way as in the case of $n=1$, if the transition $\transa$ is not 
  in the persistent set in $s'_{m+1}$, then we can delay $\transa$ as long as
  possible by taking the greatest $k > 0$ such that there is a path
  \[
      s'_{m+1} \transX{\transc_m} s'_{m+2}  \transX{\transc_{m+1}} 
     \cdots 
     \transX{\transc_{m+k-1}} s'_{m+k+1},
  \] 
  where for all $l=1,\ldots,k+1$, the set $P_{m+l}$ is the persistent set
  in state $s'_{m+l}$, the transition $\transc_{m+l-1}$ is in $P_{m+l}$, and 
  the transition $\transa$ in not in $P_{m+l}$. 
  Thus, the path
  \[
      s_0 \transX{\transc_0} s_1  \transX{\transc_1} 
     \cdots 
     \transX{\transc_{m-1}} s_m \transX{\transb_0} s'_{m+1}
     \cdots 
     \transX{\transc_{m+k-1}} s'_{m+k+1} \transX{\transa} s'_e 
  \] 
  is the path from $s_0$ leading to an error state $s'_e$ such that it 
  consists only of transitions in the persistent sets of visited states.

  \item\noindent{\hskip-12 pt\bf Case $n>1$:}\ Let $s_0 \in S$ be such that there is a path
  \[
     s_0 \transX{\transb_0} s'_1 \transX{\transb_1}  
     \cdots
     \transX{\transb_{n-1}=\transa} s_e
  \]  
  for some state $s'_1$ and an error state $s_e$. By the successor-state 
  condition, the persistent set in $s_0$ is non-empty. If the only persistent 
  set in $s_0$ is the singleton set $\setof{\transb_0}$, then the path
  $s_0 \transX{\transb_0} s'_1$ consists only of transition in the 
  persistent set. By the case $n-1$, it is guaranteed that there is a path 
  from $s'_1$ leading to an error state $s'_e$ such that the path consists
  only of transitions in the persistent sets of visited states. Thus, 
  there is a path from $s_0$ leading to an error state $s'_e$ such that 
  the path consists only of transitions in the persistent sets of visited
  states.

  Suppose that the transition $\transb_0$ is not in the persistent set 
  in $s_0$. Take the greatest $m > 0$ such that there is a path
  \[
      s_0 \transX{\transc_0} s_1  \transX{\transc_1} 
     \cdots 
     \transX{\transc_{m-1}} s_m,
  \] 
  where for all $i=0,\ldots,m-1$, the set $P_i$ is the persistent set
  in state $s_i$, the transition $\transc_i$ is in $P_i$, and the
  transition $\transb_0$ in not in $P_i$ (see 
  Figure~\ref{fig:SafetyPreserveN}). That is, we delay the exploration 
  of $\transb_0$ as long as possible. 

  Consider now the path 
  \[
      s_0 \transX{\transc_0} s_1  \transX{\transc_1} 
     \cdots 
     \transX{\transc_{m-1}} s_m \transX{\transb_0} s'_{m+1}.
  \] 
  With the same reasoning as in the case of $n=2$, we have the transition
  $\transb_1$ enabled in the state $s'_{m+1}$, and we can postpone 
  the exploration of $\transb_1$ as long as possible. When 
  $\transb_1$ gets explored, the transition $\transb_2$ is enabled in 
  the successor state. By repeatedly applying the same reasoning for 
  transitions $\transb_k$ for $k=2,\ldots,n-1$, the path formed in a similar
  way to that of the case of $n=2$ is the path from $s_0$ leading to an error
  state $s'_e$ such that the path consists only of transitions in the 
  persistent sets of visited states. \qed
\end{desCription}

\begin{figure}
\centering
  \begin{tabular}{ccc}
\includegraphics[width=0.24\textwidth]{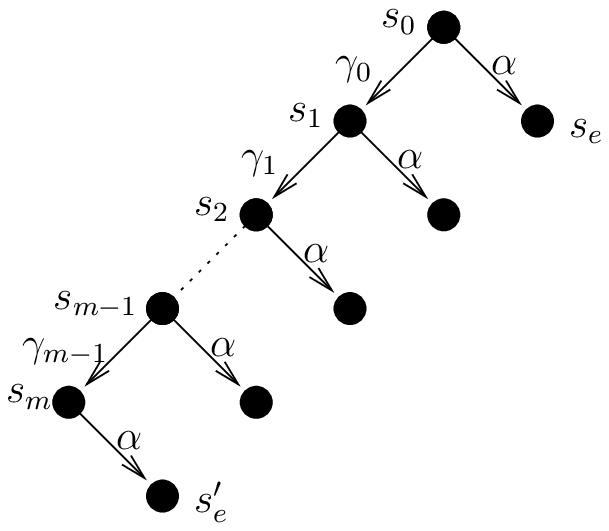} &
\includegraphics[width=0.39\textwidth]{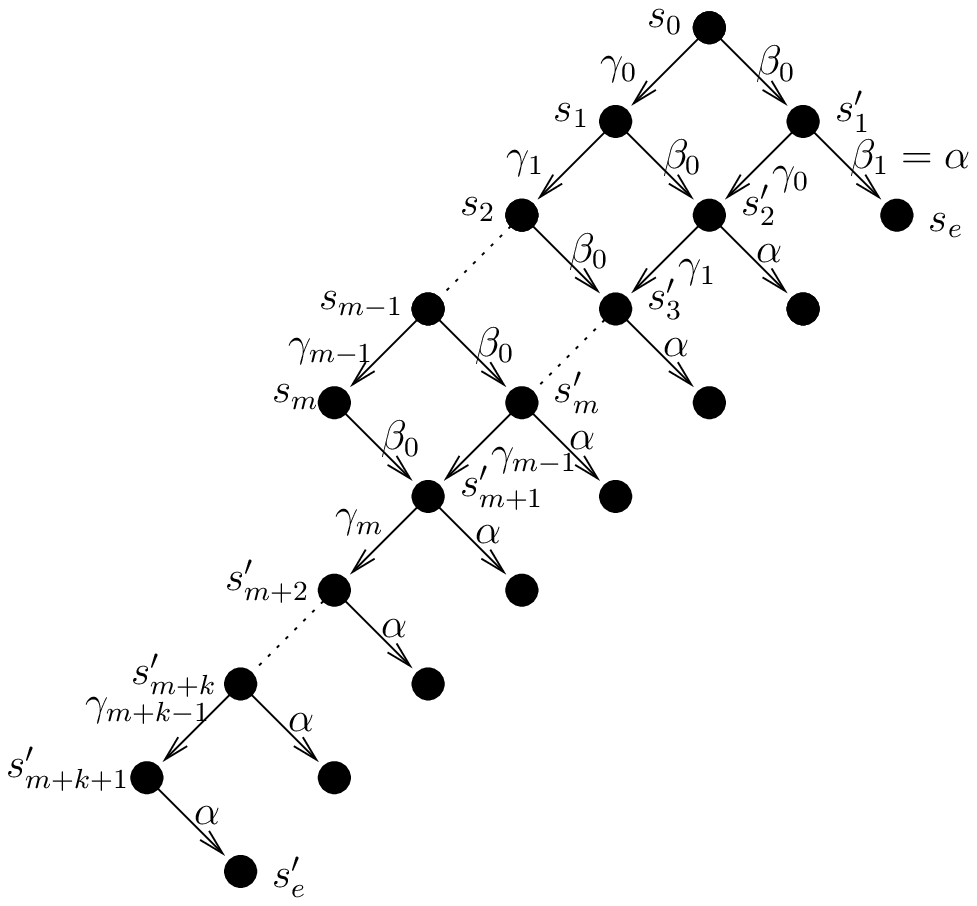} &
\includegraphics[width=0.35\textwidth]{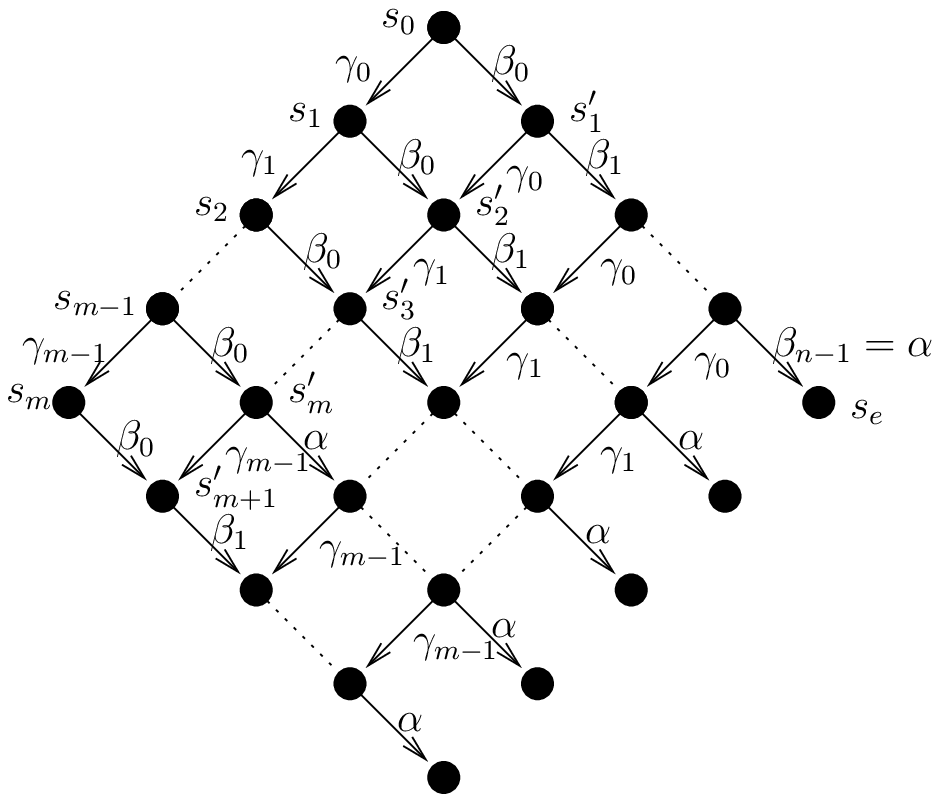} \\
 $n=1$ & $n=2$ & $n>1$
  \end{tabular}
\caption{Cases of the proof of Theorem~\ref{thm:PORCorrect}. \label{fig:SafetyPreserve1}\label{fig:SafetyPreserve2}\label{fig:SafetyPreserveN}}
\end{figure}
}

\subsubsection{The Sleep Set Approach}

The \emph{sleep set} \POR technique exploits independencies of enabled 
transitions in the current state. 
For example, suppose that in some state $s$ there are two enabled
transitions $\transa$ and $\transb$, and they are independent of each
other. 
Suppose further that the search explores $\transa$ first from $s$. 
Then, when the search explores $\transb$ from $s$ such that
$\StransbS{s}{s'}$ for some state $ s'$, we associate with $s'$ a
sleep set containing only $\transa$. 
From $s'$ the search only explores transitions that are not in the
sleep set of $s'$. That is, although the transition $\transa$ is still 
enabled in $s'$, it will not be explored. 
Both persistent set and sleep set techniques are orthogonal and
complementary, and thus can be applied simultaneously.
Note that the sleep set technique only removes transitions, and not
states. Thus, Theorem~\ref{thm:PORCorrect} still holds when the sleep set
technique is applied.

\subsection{Applying POR to ESST}
\label{subsec:ApplyPORToESST}

The key idea of applying \POR to \ESST is to select a representative
subset of scheduler states output by the scheduler in \ESST. That is,
instead of creating successor nodes with all scheduler states 
from $\setof{\sst_1,\ldots,\sst_n} = \Sched(\sst)$, for some state $\sst$, 
we create successor nodes with the representative subset of 
$\setof{\sst_1,\ldots,\sst_n}$. However, such an application is non-trivial.
The \ESST algorithm is based on the construction of an \ARF that 
describes the reachable abstract states, while the exposition of \POR before 
is based on the analysis of reachable concrete states. As we will see later, 
some \POR properties that hold in the concrete state space do not hold in 
the abstract state space. Nevertheless, in applying \POR to \ESST one needs 
to guarantee that the original \ARF is safe if and only if the reduced \ARF, 
obtained by the restriction on the scheduler's output, is safe. 
In particular, the construction of reduced \ARF has to check if the cycle 
condition is satisfied in its concretization.

To integrate \POR techniques into the \ESST algorithm, we first need to
identify fragments in the threaded program that count as transitions
in the transition system. In the previous description of \POR the execution 
of a transition is atomic, that is, its execution cannot be interleaved
by the executions of other transitions.
We introduce the notion of atomic block as the notion of transition in
the threaded program. Intuitively, an atomic block is a block of
operations between calls to primitive functions that can suspend the
thread.  
Let us call such primitive functions \emph{blocking functions}.

An \emph{atomic block} of a thread is a rooted subgraph of 
the \CFG such that the subgraph satisfies the following conditions:
\begin{enumerate}[(1)]
  \item its unique entry is the entry of the \CFG or the location that
  immediately follows a call to a blocking function; 

  \item its exit is the exit of the \CFG or the location that immediately
  follows a call to a blocking function; and 

  \item there is no call to a blocking function in any \CFG path from the
  entry to an exit except the one that precedes the exit.  
\end{enumerate}
Note that an atomic block has a unique entry, but can have multiple
exits. We often identify an atomic block by its entry. Furthermore,
we denote by $\ABlock$ the set of atomic blocks.

\begin{exa}
Consider a thread whose \CFG is depicted in
Figure~\ref{fig:AtomicBlocks}(a).
Let \code{wait($\ldots$)} be the only call to a blocking function in
the \CFG.
Figures~\ref{fig:AtomicBlocks}(b) and (c) depicts the atomic blocks of
the thread. The atomic block in Figure~\ref{fig:AtomicBlocks}(b)
starts from $l_0$ and exits at $l_5$ and $l_7$, while the one in
Figure~\ref{fig:AtomicBlocks}(c) starts from $l_5$ and exits at $l_5$
and $l_7$.

\begin{figure}
\begin{tabular}{ccc}

\begin{minipage}{0.3\textwidth}
\begin{center}
{\scriptsize
\psset{xunit=0.5cm,yunit=0.65cm,labelsep=2pt}
\begin{pspicture}(0,0)(6,8)

\rput(2,7.5){\circlenode{l0}{$l_0$}}
\rput(2,6.1){\circlenode{l1}{$l_1$}}
\rput(0.5,4.7){\circlenode{l2}{$l_2$}}
\rput(3.5,4.7){\circlenode{l3}{$l_3$}}
\rput(3.5,3.3){\circlenode{l4}{$l_4$}}
\rput(3.5,1.9){\circlenode{l5}{$l_5$}}
\rput(3.5,0.5){\circlenode{l6}{$l_6$}}
\rput(0.5,0.5){\circlenode{l7}{$l_7$}}

\ncline{->}{l0}{l1}

\ncline{->}{l1}{l2}
\nbput{$\ldots$}

\ncline{->}{l1}{l3}
\naput{$\ldots$}

\ncline{->}{l2}{l7}

\ncline[linestyle=dotted]{->}{l3}{l4}

\ncline{->}{l4}{l5}
\naput{\texttt{x:=wait($\ldots$)}}

\ncline[linestyle=dotted]{->}{l5}{l6}

\ncangles[linearc=.2,armA=1.7]{->}{l6}{l1}

\end{pspicture}
}
\end{center}
\end{minipage}
&
\begin{minipage}{0.33\textwidth}
\begin{center}
{\scriptsize
\psset{xunit=0.5cm,yunit=0.65cm,labelsep=2pt}
\begin{pspicture}(0,0)(6,8)

\rput(2,7.5){\circlenode{l0}{$l_0$}}
\rput(2,6.1){\circlenode{l1}{$l_1$}}
\rput(0.5,4.7){\circlenode{l2}{$l_2$}}
\rput(3.5,4.7){\circlenode{l3}{$l_3$}}
\rput(3.5,3.3){\circlenode{l4}{$l_4$}}
\rput(3.5,1.9){\circlenode{l5}{$l_5$}}
\rput(3.5,0.5){\circlenode{l6}{$l_6$}}
\rput(0.5,0.5){\circlenode{l7}{$l_7$}}

\ncline{->}{l0}{l1}

\ncline{->}{l1}{l2}
\nbput{$\ldots$}

\ncline{->}{l1}{l3}
\naput{$\ldots$}

\ncline{->}{l2}{l7}

\ncline[linestyle=dotted]{->}{l3}{l4}

\ncline{->}{l4}{l5}
\naput{\texttt{x:=wait($\ldots$)}}

\ncline[linestyle=dotted]{->}{l5}{l6}

\ncangles[linearc=.2,armA=1.7]{->}{l6}{l1}

\pscurve[linestyle=dashed,linecolor=red]
  (2,8)(2.5,7.9)(2.5,7)(3,6)(4,5)
  (4,2)(3.5,1.8)(3,2)(3,4)(2,5)(1,4.5)
  (1,0.8)(0.5,0.5)(0,1)(0,5)(1.3,6)
  (1.3,7.9)(2,8)

\end{pspicture}
}
\end{center}
\end{minipage}
&
\begin{minipage}{0.33\textwidth}
\begin{center}
{\scriptsize
\psset{xunit=0.5cm,yunit=0.65cm,labelsep=2pt}
\begin{pspicture}(0,0)(6,8)

\rput(2,7.5){\circlenode{l0}{$l_0$}}
\rput(2,6.1){\circlenode{l1}{$l_1$}}
\rput(0.5,4.7){\circlenode{l2}{$l_2$}}
\rput(3.5,4.7){\circlenode{l3}{$l_3$}}
\rput(3.5,3.3){\circlenode{l4}{$l_4$}}
\rput(3.5,1.9){\circlenode{l5}{$l_5$}}
\rput(3.5,0.5){\circlenode{l6}{$l_6$}}
\rput(0.5,0.5){\circlenode{l7}{$l_7$}}

\ncline{->}{l0}{l1}

\ncline{->}{l1}{l2}
\nbput{$\ldots$}

\ncline{->}{l1}{l3}
\naput{$\ldots$}

\ncline{->}{l2}{l7}

\ncline[linestyle=dotted]{->}{l3}{l4}

\ncline{->}{l4}{l5}
\naput{\texttt{x:=wait($\ldots$)}}

\ncline[linestyle=dotted]{->}{l5}{l6}

\ncangles[linearc=.2,armA=1.7]{->}{l6}{l1}

\pscurve[linestyle=dashed,linecolor=red]
  (3.5,2.3)(4,2.1)(4,0.9)(5.1,0.9)
  (5.1,5.7)(3,5.7)(4,5)(4,2)(3.5,1.8)
  (3,2)(3,4)(2,5)(1,4.5)(1,0.8)(0.5,0.5)
  (0,1)(0,5)(1.3,6.5)(5.7,6.4)(5.7,0.5)
  (5.5,0.1)(3,0.1)(3,2.1)(3.5,2.3)

\end{pspicture}
}
\end{center}
\end{minipage} \\
(a) & (b) & (c)
\end{tabular}
\caption{Identifying atomic blocks. \label{fig:AtomicBlocks}}
\end{figure}

\end{exa}

Note that, an atomic block can span over multiple basic blocks or even
multiple large blocks in the basic block or large block
encoding~\cite{fmcad2009}.
In the sequel we will use the terms transition and atomic block
interchangeably.

Prior to computing persistent sets, we need to compute valid dependence 
relations. 
The criteria for two transitions being dependent are different from
one application domain to the other. Cooperative threads in many 
embedded system domains employ event-based synchronizations through
event waits and notifications. Different domains can have different types
of event notification. For generality, we anticipate two kinds of 
notification: immediate and delayed notifications. An immediate notification 
is materialized immediately at the current time or at the current cycle 
(for cycle-based semantics). Threads that are waiting for the notified events 
are made runnable upon the notification. A delayed notification is scheduled to
be materialized at some future time or at the end of the current cycle. 
In some domains delayed notifications can be cancelled before they are 
triggered.

For example, in a system design language that supports event-based
synchronization, a pair $(\transa,\transb)$ of atomic blocks are in a
valid dependence relation if one of the following criteria is
satisfied:
(1) the atomic block $\transa$ contains a write to a shared (or
global) variable $g$, and the atomic block $\transb$ contains a write
or a read to $g$;
(2) the atomic block $\transa$ contains an immediate notification of
an event $e$, and the atomic block $\transb$ contains a wait for $e$;
(3) the atomic block $\transa$ contains a delayed notification of an
event $e$, and the atomic block $\transb$ contains a cancellation of a
notification of $e$.
Note that the first criterion is a standard criterion for two blocks 
to become dependent on each other. 
That is, the order of executions of the two blocks is relevant because
different orders yield different values assigned to variables. 
The second and the third criteria are specific to event-based
synchronization language. 
An event notification can make runnable a thread that is waiting for a
notification of the event.  
A waiting thread misses an event notification if the thread waited for
such a notification after another thread had made the notification.
Thus, the order of executions of atomic blocks containing event waits
and event notifications is relevant. 
Similarly for the delayed notification in the third criterion. 
Given criteria for being dependent, one can use static analysis
techniques to compute a valid dependence relation.

\begin{algorithm}[t]
\caption{Persistent sets.}
\label{alg:PersistentSet}
\begin{tabular}{l}
\textbf{Input}: a set $B_{en}$ of enabled atomic blocks. \\
\textbf{Output}: a persistent set $P$. \\
\begin{minipage}{\textwidth}
\begin{enumerate}
  \item Let $B := \setof{\transa}$, where $\transa \in B_{en}$.
  \item For each atomic block $\transa \in B$:
  \begin{enumerate}
    
    \item If $\transa \in B_{en}$ ($\transa$ is enabled):
      \begin{iteMize}{$\bullet$}
        \item Add into $B$ every atomic block $\transb$
         such that $(\transa,\transb) \in D$.
      \end{iteMize}
    
    \item If $\transa \not\in B_{en}$ ($\transa$ is disabled):
      \begin{iteMize}{$\bullet$}
        \item Add into $B$ a necessary enabling set for $\transa$ with 
        $B_{en}$. 
      \end{iteMize}
  \end{enumerate}
  \item Repeat step 2 until no more atomic blocks can be added into $B$.
  \item $P := B \cap B_{en}$.
\end{enumerate}
\end{minipage}
\end{tabular}
\end{algorithm}

To have small persistent sets, we need to know whether a disabled transition
that has a dependence relation with the currently enabled ones can
be made enabled in the future. To this end, we use the notion of
necessary enabling set introduced in~\cite{DBLP:books/sp/Godefroid96}.

\begin{defi}[Necessary Enabling Set]\label{def:NES}
Let $M = (S,S_0,T)$ be a transition system such that a transition 
$\transa \in T$ is diabled in a state $s \in S$. A set 
$T_{\transa,s} \subseteq T$
is a \emph{necessary enabling set for $\transa$ in $s$} if for every
finite path
$
  s = s_0 \transX{\transa_0} \cdots \transX{\transa_{n-1}} s_n 
$
in $M$ such that $\transa$ is disabled in $s_i$, for all $0 \leq i < n$,
but is enabled in $s_n$, a transition $t_j$, for some $0 \leq j \leq n-1$,
is in $T_{\transa,s}$.
A set $T_{\transa,T_{en}} \subseteq T$, for $T_{en} \subseteq T$, 
is a \emph{necessary enabling set for $\transa$ with $T_{en}$} if
$T_{\transa,T_{en}}$ is a necessary enabling set for $\transa$ in every 
state $s$ such that $T_{en}$ is the set of enabled transitions in $s$.
\end{defi}

\noindent Intuitively, a necessary enabling set $T_{\transa,s}$ for 
a transition $\transa$ in a state $s$ is a set of transitions such 
that $\transa$ cannot become enabled in the future before at least 
a transition in $T_{\transa,s}$ is executed.

Algorithm~\ref{alg:PersistentSet} computes persistent sets using 
a valid dependence relation $D$. 
It is easy to see that the persistent set computed by the algorithm
satisfies the successor-state condition.
The algorithm is also a variant of the stubborn set algorithm presented 
in~\cite{DBLP:books/sp/Godefroid96}, that is, we use a valid dependence 
relation as the interference relation used in the latter algorithm.

We apply \POR to the \ESST algorithm by modifying the \ARF node
expansion rule~\ref{rule:ARFExp2}, described in Section~\ref{sec:ESST}
in two steps. 
First we compute a persistent set from a set of scheduler states
output by the function $\Sched$.
Second, we ensure that the cycle condition is satisfied by the
concretization of the constructed \ARF.

We introduce the function $\Persistent$ that computes a persistent set of 
a set of scheduler states.
$\Persistent$ takes as inputs an \ARF node and a set $\schedstset$ of
scheduler states, and outputs a subset $\schedstset'$ of $\schedstset$.
The input \ARF node keeps track of the thread locations, which are
used to identify atomic blocks, while the input scheduler states keep
track of the status of the threads.
From the \ARF node and the set $\schedstset$, the function $\Persistent$
extracts the set $B_{en}$ of enabled atomic blocks. $\Persistent$ then
computes a persistent set $P$ from $B_{en}$ using
Algorithm~\ref{alg:PersistentSet}.
Finally, $\Persistent$ constructs back a subset $\schedstset'$ of the
input set $\schedstset$ of scheduler states from the persistent set
$P$.

Let $\frnode = (\ststate{l_1}{\varphi_1},\ldots,\ststate{l_N}{\varphi_N},
\varphi,\sst)$ be an \ARF node that is going to be expanded. 
We replace the rule~\ref{rule:ARFExp2} in the following way: instead of
creating a new \ART for each state $\sst' \in \Sched(\sst)$, we create a new
\ART whose root is the node
$(\ststate{l_1}{\varphi_1},\ldots,\ststate{l_N}{\varphi_N},
\varphi,\sst')$ for each state $\sst' \in \Persistent(\frnode,\Sched(\sst))$
(rule~\ref{rule:ARFExp2}').

To guarantee the preservation of safety properties, we have to check
that the cycle condition is satisfied.
Following~\cite{ClarkeGrumbergPeledBook99}, we check a stronger
condition: at least one state along the cycle is fully expanded.
In the \ESST algorithm a \emph{potential} cycle occurs if an \ARF node
is covered by one of its predecessors in the \ARF.
Let $\frnode = (\ststate{l_1}{\varphi_1},\ldots,\ststate{l_N}{\varphi_N},\varphi,\sst)$
be an \ARF node. We say that the scheduler state $\sst$ is \emph{running}
if there is a running thread in $\sst$. 
%
%
We also say that the node $\frnode$ is \emph{running} if its scheduler
state $\sst$ is.
Note that during \ARF expansion the input of $\Sched$ is always a
non-running scheduler state.
A path in an \ARF can be represented as a sequence
$\frnode_0,\ldots,\frnode_m$ of \ARF nodes such that for all $i$, we
have $\frnode_{i+1}$ is a successor of $\frnode_i$ in the same \ART
or there is an \ARF connector from $\frnode_i$ to $\frnode_{i+1}$.
Given an \ARF node $\frnode$ of \ARF $\forest$, we denote by
$\arfpath{\frnode}{\forest}$ the \ARF path
$\frnode_0,\ldots,\frnode_m$ such that $\frnode_0$ has neither a
predecessor \ARF node nor an incoming \ARF connector, and $\frnode_m =
\frnode$. Let $\fpath$ be an \ARF path, we denote by
$\nonrunning{\fpath}$ the \emph{maximal} subsequence of non-running
node in $\fpath$.

\begin{algorithm}[t]
\caption{\ARF expansion algorithm for non-running node.}
\label{alg:ARFExpansion}
\begin{tabular}{l}
\textbf{Input}: a non-running \ARF node $\frnode$ that contains no error locations.\\
\begin{minipage}{\textwidth}
\begin{enumerate}
  \item Let $\nonrunning{\arfpath{\frnode}{\forest}}$ be $\frnode_0,\ldots,\frnode_m$ such  that 
  $\frnode = \frnode_m$
  
  \item If there exists $i < m$ such that $\frnode_i$ covers $\frnode$:
  \begin{enumerate} 
    \item Let $\frnode_{m-1}=(\ststate{l'_1}{\varphi'_1},\ldots,
              \ststate{l'_N}{\varphi'_N},\varphi',\sst')$.
    \item If $\Persistent(\frnode_{m-1},\Sched(\sst')) \subset \Sched(\sst')$:
      \begin{iteMize}{$\bullet$}
        \item For all $\sst'' \in \Sched(\sst')
                       \setminus\Persistent(\frnode_{m-1},\Sched(\sst'))$:
          \begin{iteMize}{$-$}
            \item  Create a new \ART with root node 
                   $(\ststate{l'_1}{\varphi'_1},\ldots,
                     \ststate{l'_N}{\varphi'_N},\varphi',\sst'')$. 
          \end{iteMize}
      \end{iteMize}
  \end{enumerate}

  \item If $\frnode$ is covered: Mark $\frnode$ as covered.
  \item If $\frnode$ is not covered: Expand $\frnode$ by rule \ref{rule:ARFExp2}'.
\end{enumerate}
\end{minipage}
\end{tabular}
\end{algorithm}

Algorithm~\ref{alg:ARFExpansion} shows how a non-running \ARF node
$\frnode$ is expanded in the presence of \POR.
We assume that $\frnode$ is not an error node. The algorithm
fully expands the immediate non-running predecessor node of $\frnode$
when a potential cycle is detected. Otherwise the node is expanded as
usual.

Our \POR technique slightly differs from that 
of~\cite{ClarkeGrumbergPeledBook99}.
On computing the successor states of a state $s$, the technique 
in~\cite{ClarkeGrumbergPeledBook99} tries to compute a persistent set $P$
in $s$ that does not create a cycle. That is, particularly for the depth-first
search (DFS) exploration, for every $\transa$ in $P$, the successor 
state $\transaS{s}$ is not in the DFS stack. If it does not succeed, then 
it fully expands the state. Because the technique 
in~\cite{ClarkeGrumbergPeledBook99} is applied to the explicit-state model 
checking, computing the successor state $\transaS{s}$ is cheap.

In our context, to detect a cycle, one has to expand an \ARF node by 
a transition (or an atomic block) that can span over multiple operations
in the \CFG, and thus may require multiple applications of the 
rule~\ref{rule:ARFExp1}. As the rule involves expensive computations
of abstract strongest post-conditions, detecting a cycle using the
technique in~\cite{ClarkeGrumbergPeledBook99} is bound to be expensive.

In addition to coverage check, in the above algorithm one can also
check if the detected cycle is spurious.
We only fully expand a node iff the detected cycle is not spurious.
When cycles are rare, the benefit of \POR can be defeated by the price
of generating and solving the constraints representing the cycle.

\POR based on sleep sets can also be applied to \ESST. 
First, we extend the node of \ARF to include a sleep set. 
That is, an \ARF node is a tuple
$(\ststate{l_1}{\varphi_1},\ldots,\ststate{l_N}{\varphi_N},\varphi,\sst,Z)$, 
where the sleep set $Z$ is a set of atomic blocks.  
The sleep set is ignored during coverage check. 
Second, from the set of enabled atomic blocks and the sleep set of the
current node, we compute a subset of enabled atomic blocks and a
mapping from every atomic block in the former subset to a successor
sleep set.

Let $D$ be a valid dependence relation, Algorithm~\ref{alg:SleepSet}
shows how to compute a reduced set of enabled transitions $B_{red}$ and a
mapping $M_Z$ to successor sleep sets using $D$. 
The input of the algorithm is a set $B_{en}$ of enabled atomic blocks and
the sleep set $Z$ of the current node.
Note that the set $B_{en}$ can be a persistent set obtained by
Algorithm~\ref{alg:PersistentSet}.

\begin{algorithm}[t]
\caption{Sleep sets.}
\label{alg:SleepSet}

\begin{tabular}{l}
\textbf{Input}: \\
\begin{minipage}{\textwidth}
\begin{itemize}
  \item a set $B_{en}$ of enabled atomic blocks.
  \item a sleep set $Z$.
\end{itemize} 
\end{minipage} \\
\textbf{Output}:\\
\begin{minipage}{\textwidth}
\begin{itemize}
  \item a reduced set $B_{red} \subseteq B_{en}$ of enabled atomic blocks.
  \item a mapping $M_Z : B_{red} \rightarrow \powerset{\ABlock}$
\end{itemize}
\end{minipage} \\

\begin{minipage}{\textwidth}
\begin{enumerate}
  \item $B_{red} := B_{en} \setminus Z$.
  \item For all $\transa \in B_{red}$:
  \begin{enumerate}

    \item For all $\transb \in Z$:
       \begin{iteMize}{$\bullet$}
         \item If $(\transa,\transb) \not\in D$ ($\transa$ and $\transb$ are
               independent):
               $M_Z[\transa] := M_Z[\transa] \cup \setof{\transb}$.
       \end{iteMize}
     
    \item $Z := Z \cup \setof{\transa}$.
  \end{enumerate}
  
\end{enumerate}
\end{minipage}

\end{tabular}

\end{algorithm}

Similar to the persistent set technique, we introduce the
function $\Sleep$ that takes as inputs an \ARF node $\frnode$ and a
set of scheduler states $\schedstset$, and outputs a subset
$\schedstset'$ of $\schedstset$ along with the above mapping $M_Z$.
From the \ARF node and the scheduler states, $\Sleep$ extracts the set
$B_{en}$ of enabled atomic blocks and the current sleep set. $\Sleep$
then computes a subset $B_{red}$ of $B_{en}$ of enabled atomic blocks
and the mapping $M_Z$ using Algorithm~\ref{alg:SleepSet}.
Finally, $\Sleep$ constructs back a subset $\schedstset'$ of the input set 
$\schedstset$ of scheduler states from the set $B_{red}$ of enabled atomic
blocks.

Let $\frnode = (\ststate{l_1}{\varphi_1},\ldots,\ststate{l_N}{\varphi_N},
\varphi,\sst,Z)$ be an \ARF node that is going to be expanded. 
We replace the rule~\ref{rule:ARFExp2} in the following way: let
$(\schedstset',M_Z) = \Sleep(\frnode,\Sched(\sst))$, create a new \ART
whose root is the node
$(\ststate{l_1}{\varphi_1},\ldots,\ststate{l_N}{\varphi_N},
\varphi,\sst',M_Z[l'])$ for each $\sst' \in \schedstset'$ such that
$l'$ is the atomic block of the running thread in $\sst'$
(rule~\ref{rule:ARFExp2}'').

One can easily combine persistent and sleep sets by replacing the above
computation $(\schedstset',M_Z) = \Sleep(\frnode,\Sched(\sst))$ by 
$(\schedstset',M_Z) = \Sleep(\frnode,\Persistent(\frnode,\Sched(\sst)))$.

\subsection{Correctness of ESST + POR}
\label{subsec:CorrectESSTPlusPOR}

The correctness of \POR with respect to verifying program assertions
in transition systems has been shown in
Theorem~\ref{thm:PORCorrect}.  
The correctness proof relies on the enabledness and commutativity of
independent transitions.  
However, the proof is applied in the concrete state space of the
transition system, while the \ESST algorithm works in the abstract state
space represented by the \ARF. 
The following observation shows that two transitions that are
independent in the concrete state space may not commute in the
abstract state space.

\begin{figure}
\centering
\includegraphics[scale=0.83]{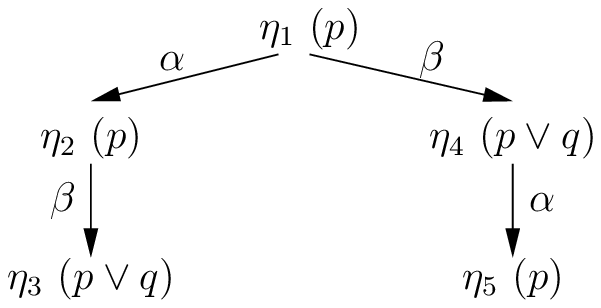}\vspace*{-5mm}
\caption{Independent transitions\newline do not commute in\newline abstract state space.
         \label{fig:non-commute}}
\end{figure}
For simplicity of presentation, we represent an abstract state by a
formula representing a region.
Let $g_1,g_2$ be global variables, and $p,q$ be predicates such that
$p \IFF (g_1 < g_2)$ and $q \IFF (g1 = g2)$.  
Let $\transa$ be the transition \code{$g_1$ := $g_1$ - 1} and
$\transb$ be the transition \code{$g_2$ := $g_2$ - 1}. 
It is obvious that $\transa$ and $\transb$ are independent of each
other. 
However, Figure~\ref{fig:non-commute} shows that the two transitions
do no commute when we start from an abstract state $\frnode_1$ such
that $\frnode_1 \IFF p$.
The edges in the figure represent the computation of abstract
strongest post-condition of the corresponding abstract states and
transitions.

Even though two independent transitions do not commute in the abstract
state space, they still commute in the concrete state space
overapproximated by the abstract state space, as shown by the lemma
below.

\begin{lem}\label{lem:CommutePreserve}
Let $\transa$ and $\transb$ be transitions that are independent of
each other such that for concrete states $s_1,s_2,s_3$ and abstract
state $\frnode$ we have $s_1 \models \frnode$, and both
$\transaSS{s_1}{s_2}$ and $\transbSS{s_2}{s_3}$ hold.
Let $\frnode'$ be the abstract successor state of $\frnode$ by
applying the abstract strongest post-operator to $\frnode$ and
$\transb$, and $\frnode''$ be the abstract successor state of
$\frnode'$ by applying the abstract strongest post-operator to
$\frnode'$ and $\transa$.
Then, there are concrete states $s_4$ and $s_5$ such that:
$\transbSS{s_1}{s_4}$ holds, 
$s_4 \models \frnode'$, 
$\transbSS{s_4}{s_5}$ holds, 
$s_5 \models \frnode''$, and 
$s_3 = s_5$.
\end{lem}
\newcommand{\ProofLemCommutePreserve}{
\proof
By the independence of $\transa$ and $\transb$, we have $\transbSS{s_1}{s_4}$
holds. By the abstract strongest post-operator, we have $s_4 \models \frnode'$.
By the independence of $\transa$ and $\transb$, we have $\transbSS{s_4}{s_5}$
holds. By the abstract strongest post-operator and the fact that 
$s_4 \models \frnode'$, we have $s_5 \models \frnode''$. 
Finally by the independence of $\transa$ and $\transb$, we have $s_3 = s_5$. 
\qed
}

The above lemma shows that \POR can be applied in the abstract state space.
Let $\ESST_{\POR}$ be the \ESST algorithm with \POR. 
The correctness of \POR in \ESST is stated by the following theorem:

\begin{thm}\label{thm:CorrectPORESST} 
Let $P$ be a threaded sequential program. For every
terminating executions of $\ESST(P)$ and $\ESST_{\POR}(P)$, we have
that $\ESST(P)$ reports safe iff so does $\ESST_{\POR}(P)$.
\end{thm} 
\newcommand{\ProofThmPORESSTCorrect}{
\proof
First, we first prove the left-to-right direction of iff and then
prove the other direction.

  $\mathbf{(\Longrightarrow):}$ Assume that $\ESST(P)$ returns 
  a safe \ARF $\forest$.
  Assume to the contrary that $\ESST_{POR}$ reports unsafe and returns
  a counter-example path $\fpath$. By Theorem~\ref{thm:CorrectESST},
  we have $\cg \simtrans{\fpath} \cg'$ for an initial configuration
  $\cg$ and an error configuration $\cg'$ of $P$. That is, the error
  configuration is in $\ReachC{P}$. 
  Again, by Theorem~\ref{thm:CorrectESST}, there is an \ARF node
  $\frnode \in \NodesF{\forest}$ such that $\cg' \models \frnode$.
  But then the node $\frnode$ is an error node, and $\forest$ is not
  safe, which contradicts our assumption that $\forest$ is safe.

  $\mathbf{(\Longleftarrow):}$ We lift Theorem~\ref{thm:PORCorrect}
  and its proof to the case of abstract transition system or abstract
  state space with the help of Lemma~\ref{lem:CommutePreserve}. 
  The lifting amounts to establishing correspondences between the
  transition system $M=(S,S_0,T)$ in Theorem~\ref{thm:PORCorrect} and
  the \ARF constructed by $\ESST$ and $\ESST_{POR}$. 
  First, since the executions are terminating, the set of reachable
  scheduler states is finite.  
  Now let the set of \ARF nodes reachable by the
  rules~\ref{rule:ARFExp1} and ~\ref{rule:ARFExp2} correspond to the
  set $S$. 
  That is, the set $S$ is now the set of \ARF nodes. 
  The set $S_0$ contains only the initial node.
  A transition in $T$ represents either an \ART path $\tpath$ that
  starts from the root of the \ART and ends with a leaf of the \ART,
  or an \ARF connector.
  The error transitions $T_{err}$ contains every transition in $T$
  such that the transition represents an \ART path $\tpath$ with an
  error node as the end node. The set $E_{M,T_{err}}$ consists of
  error nodes.
  Every path 
  $
      s_0 \transX{\transa_0} s_1  \transX{\transa_1} 
     \cdots 
     \transX{\transa_{n-1}} s_n,
  $ 
  corresponds the the following path in the \ARF: 
  \begin{enumerate}[(1)]
    \item for $i=0,\ldots,n$, the node $s_i$ is a node in the \ARF, 
    \item for $i=0,\ldots,n-1$, there is an \ARF path from 
    $s_i$ to $s_{i+1}$ that is represented by the transition $\transa_i$, and
    \item for $i=0,\ldots,n-1$, if the transition $\transa_i$ leads to
    a node $s$ covered by another node $s'$, then $s_{i+1}$ is $s'$.
  \end{enumerate}

  \noindent We now exemplify how we address the issue of commutativity
  in the proof of Theorem~\ref{thm:PORCorrect}. Consider the case
  $n=2$ where transitions $\transc_0,\transb_0$ and
  $\transb_0,\transc_0$ commute in $s'_2$.  In the case of abstract
  state space, they might not commute. However, by
  Lemma~\ref{lem:CommutePreserve}, they commute in the concrete state
  space. Thus, the transition $\transa$ is still enabled after
  performing the transitions $\transc_0,\transb_0$.  \qed }

%% file: experiments.tex
\section{Experimental Evaluation}
\label{sec:Application}

In this section we show an experimental evaluation of the \ESST
algorithm in the verification of multi-threaded programs in the
\fairthreads~\cite{DBLP:journals/concurrency/Boussinot06} programming
framework.
The aim of this evaluation is to show the effectiveness of \ESST and
of the partial-order reduction applied to \ESST. By following the same
methodology, the \ESST algorithm can be adapted to other programming 
frameworks, like \specc~\cite{GerstlauerEtAl:SystemDesign:01} and 
\osekvdx~\cite{OSEK}, with moderate effort.

\subsection{Verifying \fairthreads}
\label{subsec:VerifyFairThreads}

\fairthreads is a framework for programming multi-threaded software
that allows for mixing both cooperative and preemptive threads. As we want
to apply \ESST, we only deal with the cooperative threads.
\fairthreads includes a scheduler that executes threads according to 
a simple round-robin policy. 
\fairthreads also provides a programming interface that allows threads 
to synchronize and communicate with each others. Examples of synchronization
primitives of \fairthreads are as follows: \code{await($e$)} for waiting 
for the notification of event $e$ if such a notification does not exist, 
\code{generate($e$)} for generating a notification of event $e$, 
\code{cooperate} for yielding the control back to the scheduler, and 
\code{join($t$)} for waiting for the termination of thread $t$.

\begin{figure}
\centerline{\includegraphics[scale=0.7]{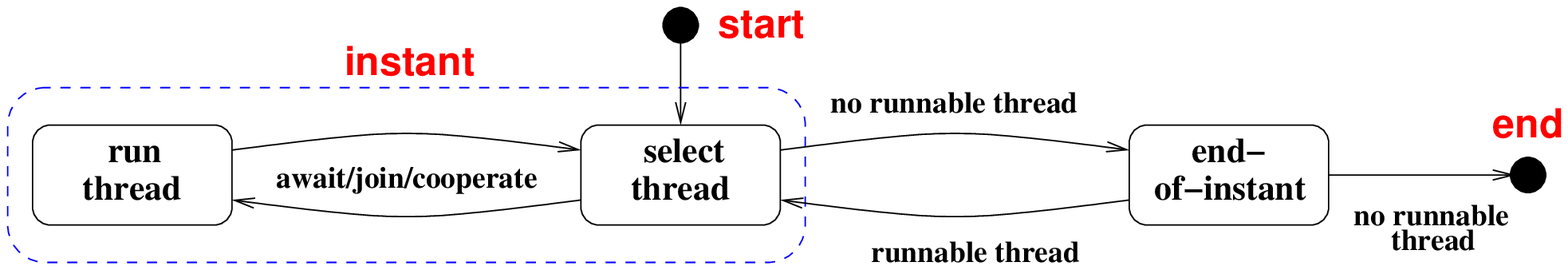}}
\caption{The scheduler of \fairthreads.\label{fig:FTScheduler}}
\end{figure}

The scheduler of \fairthreads is shown in Figure~\ref{fig:FTScheduler}.  
At the beginning all threads are set to be runnable.
The executions of threads consist of a series of \emph{instants} in
which the scheduler runs all runnable threads, in a deterministic
round-robin fashion, until there are no more runnable threads.

A running thread can yield the control back to the scheduler either 
by waiting for an event notification (\code{await}), 
by cooperating (\code{cooperate}), or by waiting for another thread 
to terminate (\code{join}). 
A thread that executes the primitive \code{await($e$)} can observe the
notification of $e$ even though the notification occurs long before
the execution of the primitive, so long as the execution of
\code{await($e$)} is still in the same instant of the notification of
$e$. Thus, the execution of \code{await} does not necessarily yields
the control back to the scheduler.

When there are no more runnable threads, the scheduler enters the
end-of-instant phase. In this phase the scheduler wakes up all threads
that had cooperated during the last instant,
and also clears all event notifications.
The scheduler then starts a new instant if there are runnable threads; 
otherwise the execution ends.

The operational semantics of cooperative \fairthreads has been
described in~\cite{Boussinot02}. 
However, it is not clear from the semantics whether the round-robin
order of the thread executions remains the same from one instant to
the other. 
Here, we assume that the order is the same from one instant to the
other. 
The operational semantics does not specify either the initial
round-robin order of the thread executions.  
Thus, for the verification, one needs to explore all possible
round-robin orders. 
This situation could easily degrade the performance of \ESST and
possibly lead to state explosion. 
The \POR techniques described in Section~\ref{sec:ESSTPlusPOR} could in
principle address this problem.

In this section we evaluate two software model checking approaches for
the verification of \fairthreads programs. In the first approach we
rely on a translation from \fairthreads into sequential programs (or
\emph{sequentialization}), such that the resulting sequential programs
contain both the mapping of the cooperative threads in the form of
functions and the encoding of the \fairthreads scheduler. The thread
activations are encoded as function calls from the scheduler function
to the functions that correspond to the threads.  The program can be
thought of as jumping back and forth between the ``control level''
imposed by the scheduler, and the ``logical level'' implemented by the
threads. Having the sequential program, we then use off-the-shelf
software model checkers to verify the programs.

In the second approach we apply the \ESST algorithm to verify \fairthreads
programs. In this approach we define a set of primitive functions that 
implement \fairthreads synchronization functions, and instantiate the
scheduler of \ESST with the \fairthreads scheduler. We then translate 
the \fairthreads program into a threaded program such that there is 
a one-to-one correspondence between the threads in the \fairthreads program 
and in the resulting threaded program. Furthermore, each call to 
a \fairthreads synchronization function is translated into a call to
the corresponding primitive function. The \ESST algorithm is then applied
to the resulting threaded program.

\subsection{Experimental evaluation setup}
\label{subsec:ExperimentsSetup}

The \ESST algorithm has been implemented in the
\kratos software model checker~\cite{DBLP:conf/cav/CimattiGMNR11}.
In this work we have extended \kratos with the \fairthreads scheduler and 
the primitive functions that correspond to the  \fairthreads synchronization
functions.

We have carried out a significant experimental evaluation on a set of 
benchmarks taken and adapted from the literature on verification 
of cooperative threads.
For example, the \code{fact*} benchmarks are extracted 
from~\cite{JohnsonBesnardGautierTalpin:AVoCS2010}, which 
describes a synchronous approach to verifying the absence of deadlocks
in \fairthreads programs.
We adapted the benchmarks by recoding the bad synchronization, that can
cause deadlocks, as an unreachable false assertion.
The \code{gear-box} benchmark is taken from the case study 
in~\cite{DBLP:journals/rts/WaszniowskiH08}.
This case study is about an automated gearbox control system that
consists of a five-speed gearbox and a dry clutch.
Our adaptation of this benchmark does not model the timing behavior of 
the components and gives the same priority to all tasks (or threads) 
of the control system. 
In our case we considered the verification of safety properties that
do not depend on the timing behavior. Ignoring the timing behavior in
this case results in more non-determinism than that of the original
case study.
The \code{ft-pc-sfifo*} and \code{ft-token-ring*} benchmarks are
taken and adapted from, respectively, the \code{pc-sfifo*} and 
\code{token-ring*} benchmarks used 
in~\cite{DBLP:conf/fmcad/CimattiMNR10,DBLP:conf/tacas/CimattiNR11}.
All considered benchmarks satisfy the restriction of \ESST:
the arguments passed to every call to a primitive function are 
constants.

For the sequentialized version of \fairthreads programs, we experimented
with several state-of-the-art predicate-abstraction based software model 
checkers, including
\satabs-3.0~\cite{DBLP:conf/tacas/ClarkeKSY05}, 
\cpachecker~\cite{DBLP:conf/cav/BeyerK11}, and 
the sequential analysis of \kratos~\cite{DBLP:conf/cav/CimattiGMNR11}.
\ignore{
We did not experiment with \blast~\cite{DBLP:journals/sttt/BeyerHJM07}
because on all benchmarks \blast results in generating error
exceptions. The new \blast-2.7 produces unsound results.
}
We also experimented with \cbmc-4.0~\cite{DBLP:conf/tacas/ClarkeKL04} 
for bug hunting with bounded model checking 
(BMC)~\cite{DBLP:conf/tacas/BiereCCZ99}. For the BMC experiment, we set
the size of loop unwindings to 5 and consider only the unsafe benchmarks.
All benchmarks and tools' setup are available at {\small
\url{http://es.fbk.eu/people/roveri/tests/jlmcs-esst}}.

We ran the experiments on a Linux machine with Intel-Xeon DC 3GHz processor
and 4GB of RAM. We fixed the time limit to 1000 seconds, and the memory limit 
to 4GB.

\subsection{Results of Experiments}
\label{subsec:ExperimentsResults}

The results of experiments are shown in Table~\ref{tab:ExpResult}, for
the run times, and in Table~\ref{tab:ExpResultPOR}, for the numbers of 
explored abstract states by \ESST.
The column V indicates the status of the benchmarks: S for safe and U
for unsafe. 
In the experiments we also enable the \POR techniques in \ESST. 
The column No-\POR indicates that during the experiments \POR is not
enabled. 
The column P-\POR indicates that only the persistent set technique is
enabled, while the column S-\POR indicates that only the sleep set
technique is enabled. 
The column PS-\POR indicates that both the persistent set and the
sleep set techniques are enabled. 
We mark the best results with bold letters, and denote the out-of-time
results by T.O.

The results clearly show that \ESST outperforms the predicate
abstraction based sequentialization approach. 
The main bottleneck in the latter approach is the number of predicates 
that the model checkers need to keep track of to model details of 
the scheduler. For example, on the \code{ft-pc-sfifo1.c} benchmark \satabs, 
\cpachecker, and the sequential analysis of \kratos needs to keep track of, 
respectively, 71, 37, and 45 predicates. On the other hand, \ESST only needs 
to keep track of 8 predicates on the same benchmark. 

Regarding the refinement steps, \ESST needs less abstraction-refinement 
iterations than other techniques. For example, starting with the empty
precision, the sequential analysis of \kratos needs 8 abstraction-refinement 
iterations to verify \code{fact2}, and 35 abstraction-refinement iterations
to verify \code{ft-pc-sfifo1}. \ESST, on the other hand, verifies \code{fact2} 
without performing any refinements at all, and verifies \code{ft-pc-sfifo1} 
with only 3 abstraction-refinement iterations.

The BMC approach, represented by \cbmc, is ineffective on our benchmarks.
First, the breadth-first nature of the BMC approach creates big formulas 
on which the satisfiability problems are hard. In particular, \cbmc employs 
bit-precise semantics, which contributes to the hardness of the problems.
Second, for our benchmarks, it is not feasible to identify the size of 
loop unwindings that is sufficient for finding the bug. For example, due
to insufficient loop unwindings, \cbmc reports safe
for the unsafe benchmarks \code{ft-token-ring-bug.4} and 
\code{ft-token-ring-bug.5} (marked with ``*''). Increasing the size of 
loop unwindings only results in time out.

Table~\ref{tab:ExpResult} also shows that the \POR techniques boost 
the performance of \ESST and allow us to verify benchmarks that could not 
be verified given the resource limits. In particular we get the best results
when the persistent set and sleep set techniques are applied together. 
Additionally, Table~\ref{tab:ExpResultPOR} shows that the \POR techniques 
reduce the number of abstract states explored by \ESST. This reduction
also implies the reductions on the number of abstract post computations 
and on the number of coverage checks.

\begin{table}[t]
\begin{center}\scriptsize
\begin{tabular}{|l|c|r|r|r|r|r|r|r|r|} \cline{3-10}
\multicolumn{2}{c|}{\ } & \multicolumn{4}{c|}{Sequentialization} & \multicolumn{4}{c|}{\ESST} \\ \hline
Name & V & \multicolumn{1}{c|}{\satabs}  &  \multicolumn{1}{c|}{\cpachecker}
         & \multicolumn{1}{c|}{\kratos}  &  \multicolumn{1}{c|}{\cbmc} 
         & \multicolumn{1}{c|}{No-\POR}
         & \multicolumn{1}{c|}{P-\POR}   &  \multicolumn{1}{c|}{S-\POR}
         & \multicolumn{1}{c|}{PS-\POR}\\\hline\hline

fact1                 & S &   9.07 & 14.26 &  2.90 &     -    & \best{0.01} & \best{0.01} & \best{0.01} & \best{   0.01 } \\\hline
fact1-bug             & U &  22.18 &  8.06 &  0.39 &   15.09  & \best{0.01} & \best{0.01} & \best{0.01} & \best{   0.03 } \\\hline
fact1-mod             & S &   4.41 &  8.18 &  0.50 &     -    &       0.40  &       0.40  & \best{0.39} & \best{   0.39 } \\\hline
fact2                 & S &  69.05 & 17.25 & 15.40 &     -    & \best{0.01} & \best{0.01} & \best{0.01} & \best{   0.01 } \\\hline
gear-box              & S &   T.O  &  T.O  &  T.O  &     -    &       T.O   &     473.55  &      44.89  & \best{  44.19 } \\\hline
ft-pc-sfifo1          & S &  57.08 & 56.56 & 44.49 &     -    &       0.30  &       0.30  & \best{0.29} & \best{   0.29 } \\\hline
ft-pc-sfifo2          & S & 715.31 &  T.O  &  T.O  &     -    &       0.39  &       0.39  & \best{0.30} & \best{   0.39 } \\\hline
ft-token-ring.3       & S & 115.66 &  T.O  &  T.O  &     -    &       0.48  &       0.29  & \best{0.20} & \best{   0.20 } \\\hline
ft-token-ring.4       & S & 448.86 &  T.O  &  T.O  &     -    &       5.20  &       1.10  & \best{0.29} & \best{   0.29 } \\\hline
ft-token-ring.5       & S &   T.O  &  T.O  &  T.O  &     -    &     213.37  &       6.20  &       0.50  & \best{   0.40 } \\\hline
ft-token-ring.6       & S &   T.O  &  T.O  &  T.O  &     -    &       T.O   &      92.39  &       0.69  & \best{   0.49 } \\\hline
ft-token-ring.7       & S &   T.O  &  T.O  &  T.O  &     -    &       T.O   &       T.O   &       0.99  & \best{   0.80 } \\\hline
ft-token-ring.8       & S &   T.O  &  T.O  &  T.O  &     -    &       T.O   &       T.O   &       1.80  & \best{   0.89 } \\\hline
ft-token-ring.9       & S &   T.O  &  T.O  &  T.O  &     -    &       T.O   &       T.O   &       3.89  & \best{   1.70 } \\\hline
ft-token-ring.10      & S &   T.O  &  T.O  &  T.O  &     -    &       T.O   &       T.O   &       9.60  & \best{   2.10 } \\\hline
ft-token-ring-bug.3   & U & 111.10 &  T.O  &  T.O  &  158.76  & \best{0.10} & \best{0.10} & \best{0.10} & \best{   0.10 } \\\hline
ft-token-ring-bug.4   & U & 306.41 &  T.O  &  T.O  & *407.36  &       1.70  &       0.30  & \best{0.10} & \best{   0.10 } \\\hline
ft-token-ring-bug.5   & U & 860.29 &  T.O  &  T.O  & *751.44  &      66.09  &       1.80  & \best{0.10} & \best{   0.10 } \\\hline
ft-token-ring-bug.6   & U &   T.O  &  T.O  &  T.O  &    T.O   &       T.O   &      26.29  &       0.20  & \best{   0.10 } \\\hline
ft-token-ring-bug.7   & U &   T.O  &  T.O  &  T.O  &    T.O   &       T.O   &       T.O   &       0.30  & \best{   0.20 } \\\hline
ft-token-ring-bug.8   & U &   T.O  &  T.O  &  T.O  &    T.O   &       T.O   &       T.O   &       0.60  & \best{   0.29 } \\\hline
ft-token-ring-bug.9   & U &   T.O  &  T.O  &  T.O  &    T.O   &       T.O   &       T.O   &       1.40  & \best{   0.60 } \\\hline
ft-token-ring-bug.10  & U &   T.O  &  T.O  &  T.O  &    T.O   &       T.O   &       T.O   &       3.60  & \best{   0.79 } \\\hline
\multicolumn{9}{c}{\ }\\[-4pt]															 
\end{tabular}
\caption{Run time results of the experimental evaluation (in seconds). \label{tab:ExpResult}}
\end{center}
\end{table}

\begin{table}[t]
\begin{center}\scriptsize
\begin{tabular}{|l|r|r|r|r|} \hline
Name & No-\POR & P-\POR & S-\POR & PS-\POR \\ \hline\hline

fact1                 &  \best{ 66} &  \best{ 66} & \best{  66} & \best{   66} \\\hline
fact1-bug             &  \best{ 49} &  \best{ 49} & \best{  49} & \best{   49} \\\hline
fact1-mod             &  \best{269} &  \best{269} & \best{ 269} & \best{  269} \\\hline
fact2                 &         49  &         49  & \best{  29} & \best{   29} \\\hline
gear-box              &        -    &     204178  &      60823  & \best{58846} \\\hline
ft-pc-sfifo1          &  \best{180} &  \best{180} & \best{ 180} & \best{  180} \\\hline
ft-pc-sfifo2          &        540  &        287  &        310  & \best{  287} \\\hline
ft-token-ring.3       &       1304  &        575  &        228  & \best{  180} \\\hline
ft-token-ring.4       &       7464  &       2483  &        375  & \best{  266} \\\hline
ft-token-ring.5       &      50364  &       7880  &        699  & \best{  395} \\\hline
ft-token-ring.6       &        -    &      32578  &       1239  & \best{  518} \\\hline
ft-token-ring.7       &        -    &        -    &       2195  & \best{  963} \\\hline
ft-token-ring.8       &        -    &        -    &       4290  & \best{ 1088} \\\hline
ft-token-ring.9       &        -    &        -    &       8863  & \best{ 2628} \\\hline
ft-token-ring.10      &        -    &        -    &      16109  & \best{ 3292} \\\hline
ft-token-ring-bug.3   &        496  &        223  &        113  & \best{   89} \\\hline
ft-token-ring-bug.4   &       2698  &        914  &        179  & \best{  125} \\\hline
ft-token-ring-bug.5   &      17428  &       2801  &        328  & \best{  181} \\\hline
ft-token-ring-bug.6   &        -    &      11302  &        611  & \best{  251} \\\hline
ft-token-ring-bug.7   &        -    &        -    &       1064  & \best{  457} \\\hline
ft-token-ring-bug.8   &        -    &        -    &       2133  & \best{  533} \\\hline
ft-token-ring-bug.9   &        -    &        -    &       4310  & \best{ 1281} \\\hline
ft-token-ring-bug.10  &        -    &        -    &       8039  & \best{ 1632} \\\hline
\end{tabular}
\caption{Numbers of explored abstract states. \label{tab:ExpResultPOR}}
\end{center}
\end{table}

Despite the effectiveness showed by the obtained results, the
following remarks are in order.
\POR, in principle, could interact negatively with the \ESST
algorithm. 
The construction of \ARF in \ESST is sensitive to the explored
scheduler states and to the tracked predicates.  
\POR prunes some scheduler states that \ESST has to explore.
However, exploring such scheduler states can yield a smaller \ARF than
if they are omitted. 
In particular, for an unsafe benchmark, exploring omitted scheduler
states can lead to the shortest counter-example path.
Furthermore, exploring the omitted scheduler states could lead to
spurious counter-example \ARF paths that yield predicates that allow
\ESST to perform less refinements and construct a smaller \ARF.

\subsection{Verifying \systemc}
\label{subsec:VerifySystemC}

\systemc is a C++ library that has widely been used to write
executable models of systems-on-chips. The library consists of a
language to model the component architecture of the system and also to
model the parallel behavior of the system by means of sequential
threads. Similar to \fairthreads, the \systemc scheduler employs a
cooperative scheduling, and the execution of the scheduler is divided
into a series of so-called delta cycles, which correspond to the
notion of instant.

Despite their similarities, the scheduling policy and the behavior of
synchronization primitives of \systemc and \fairthreads have
significant differences. For example, the \fairthreads scheduler
employs a round-robin scheduling, while the \systemc scheduler can
execute any runnable thread. Also, in \fairthreads a notification of
an event performed by some thread can later still be observed by
another thread, as long as the execution of the other thread is still
in the same instant as the notification.  In \systemc the latter
thread will simply miss the notification.

In~\cite{DBLP:conf/fmcad/CimattiMNR10,DBLP:conf/tacas/CimattiNR11}, we
report on the application of \ESST to the verification of \systemc
models. We follow a similar approach, comparing \ESST and the
sequentialization approach, and also experimenting with \POR in \ESST.
The results of those experiments show the same patterns as the results
reported here for \fairthreads: the \ESST approach outperforms the
sequentialization approach, and the \POR techniques improve further
the performance of \ESST in terms of run time and the the number of
visited abstract states. 
These results allow us to conclude that the \ESST algorithm, along
with the \POR techniques, is a very effective and general technique
for the verification of cooperative threads.

%% file: related-work.tex
\section{Related Work}
\label{sec:RelatedWork}

There have been a plethora of works on developing techniques for
the verification of multi-threaded programs, both for general ones
and for those with specific scheduling policies. Similar to the work
in this paper, many of these existing techniques are concerned with
verifying safety properties. In this section we review some of these
techniques and describe how they are related to our work.

\subsection{Verification of Cooperative Threads}
\label{subsec:CooperativeThread}

Techniques for verifying multi-threaded programs with cooperative
scheduling policy have been considered in different application 
domains:
\cite{DBLP:conf/acsd/MoyMM05,DBLP:conf/iscas/GrosseD05,DBLP:conf/memocode/KroeningS05,DBLP:conf/spin/TraulsenCMM07,DBLP:conf/codes/HerberFG08,DBLP:conf/iccad/BlancK08,DBLP:conf/fmcad/CimattiMNR10}
for \systemc, \cite{JohnsonBesnardGautierTalpin:AVoCS2010} for
\fairthreads, \cite{DBLP:journals/rts/WaszniowskiH08} for \osekvdx,
and \cite{DBLP:journals/fmsd/ClarkeJK07} for \specc.
Most of these techniques either embed details of the scheduler in the
programs under verification or simply abstract away those details.  
As shown in~\cite{DBLP:conf/fmcad/CimattiMNR10}, verification
techniques that embed details of the scheduler show poor scalability.
On the other hand, abstracting away the scheduler not only
makes the techniques report too many false positives, but also
limits their applicability.
The techniques described in~\cite{DBLP:conf/acsd/MoyMM05,%
DBLP:conf/spin/TraulsenCMM07,%
DBLP:conf/codes/HerberFG08} only employs explicit-state model checking
techniques, and thus they cannot handle effectively infinite-domain
inputs for threads. 
\ESST addresses these issues by analyzing the
threads symbolically and by orchestrating the overall verification by
direct execution of the scheduler that can be modeled faithfully.

\subsection{Thread-modular Model Checking}
\label{subsec:ThreadModularMC}

In the traditional verification methods, such as the one described
in~\cite{DBLP:journals/acta/OwickiG76}, safety properties are proved with
the help of assertions that annotate program statements.
These annotations form the pre- and post-conditions for the statements.
The correctness of the assertions is then proved by proof rules that
are similar to the Floyd-Hoare proof
rules~\cite{Floyd:AMS:1967,DBLP:journals/cacm/Hoare83} for sequential
programs.
The method in~\cite{DBLP:journals/acta/OwickiG76} requires a so-called
interference freedom test to ensure that no assertions used in the
proof of one thread are invalidated by the execution of another
thread. Such a freedom test makes this method non-modular (each thread
cannot be verified in isolation from other threads).

Jones~\cite{DBLP:journals/toplas/Jones83} introduces thread-modular
reasoning that verifies each thread separately using assumption about 
the other threads.
In this work the interference information is incorporated into the
specifications as environment assumptions and guarantee relations. 
The environment assumptions model the interleaved transitions of other
threads by describing their possible updates of shared variables. The
guarantee relations describe the global state updates of the whole
program. 
However, the formulation of the environment assumptions
in~\cite{DBLP:journals/toplas/Jones83} and
\cite{DBLP:journals/acta/OwickiG76} incurs a significant verification
cost.

Flanagan and Qadeer~\cite{DBLP:conf/spin/FlanaganQ03} describe
a thread-modular model checking technique that automatically infers
environment assumptions. 
First, a guarantee relation for each of the thread is inferred. The
assumption relation for a thread is then the disjunction of all the
guarantee relations of the other threads.
Similar to \ESST, this technique computes an over-approximation of the
reachable concrete states of the multi-threaded program by abstraction
using the environment assumptions. However, unlike \ESST, the
thread-modular model checking technique is incomplete since it can
report false positives.

Similar to \ESST, the work in~\cite{DBLP:conf/cav/HenzingerJMQ03}
describes a \CEGAR-based thread-modular model checking technique, that
analyzes the data-flow of each thread symbolically using predicate
abstraction, starting from a very coarse over-approximation of the
thread's data states and successively refining the approximation using
predicates discovered during the \CEGAR loop.
Unlike \ESST, the thread-modular algorithm also analyzes the
environment assumption symbolically starting with an empty environment
assumption and subsequently weakening it using the refined
abstractions of threads' data states.

Chaki et. al.~\cite{DBLP:journals/entcs/ChakiOYC03} describe another
\CEGAR-based model checking technique.
Like \ESST, the programs considered by this technique have a fixed number 
of threads. But, unlike other previous techniques that deal with 
shared-variable multi-threaded programs, the threads considered by 
this technique use message passing as the synchronization mechanism.
This technique uses two levels of abstractions over each individual
thread. The first abstraction level is predicate abstraction. The
second one, which is applied to the result of the first abstraction,
is action-guided abstraction. The parallel composition of the threads
is performed after the second abstraction has been applied.  
Compositional reasoning is used during the check for spuriousness of a
counter-example by projecting and examining the counter-example on
each individual thread separately.

Recently, Gupta et. al.~\cite{DBLP:conf/popl/GuptaPR11} have proposed 
a new predicate abstraction and refinement technique for verifying 
multi-threaded programs
Similar to \ESST, the technique constructs an \ART for each
thread. But unlike \ESST, branches in the constructed \ART might not
correspond to a \CFG unwinding but correspond to transitions of the
environment.
The technique uses a declarative formulation of the refinement
to describe constraints on the desired predicates for thread
reachability and environment transition.
Depending on the declarative formulation, the technique can generate 
a non-modular proof as in~\cite{DBLP:journals/acta/OwickiG76} or 
a modular proof as in~\cite{DBLP:conf/spin/FlanaganQ03}.

\subsection{Bounded Model Checking}
\label{subsec:BoundedMC}

Another approach to verifying multi-threaded programs is by bounded
model checking (BMC)~\cite{DBLP:conf/tacas/BiereCCZ99}. 
For multi-threaded programs, the bound is concerned, not only with the
length (or depth) of \CFG unwinding, as in the case of sequential
programs, but also with the number of scheduler invocations or the
number of context switches. This approach is sound and complete, but
only up to the given bound. 

Prominent techniques that exploit the BMC approach 
include~\cite{DBLP:journals/fmsd/Godefroid05}
and~\cite{DBLP:conf/tacas/QadeerR05}.
The work in~\cite{DBLP:journals/fmsd/Godefroid05} limits the number of
scheduler invocations. 
While the work in~\cite{DBLP:conf/tacas/QadeerR05} bounds the number 
of context switches. That is, given a bound $k$, the technique verifies if 
a multi-threaded program can fail an assertion through an execution with 
at most $k$ context switches. 
This technique relies on regular push-down systems~\cite{Schwoon2000} for 
a finite representation of the unbounded number of stack configurations.
The \ESST algorithm can easily be made depth bounded or context-switch
bounded by not expanding the constructed \ARF node when the number 
of \ARF connectors leading to the node has reached the bound.

The above depth bounded and context-switch bounded model checking
techniques are ineffective in finding errors that appear only after
each thread has a chance to complete its
execution. 
To overcome this limitation, 
Musuvathi and Qadeer~\cite{DBLP:conf/pldi/MusuvathiQ07} have
proposed a BMC technique that bounds the number of context
switches caused only by scheduler preemptions. Such a bound gives the
opportunity for each thread to complete its execution.

The state-space complexity imposed by the previously described BMC
techniques grows with the number of threads.
Thus, those techniques are ineffective for verifying multi-threaded
programs that allow for dynamic creations of threads.  
Recently a technique called delay bounded scheduling has been 
proposed in~\cite{DBLP:conf/popl/EmmiQR11}. Given a bound $k$, a
deterministic scheduler is made non-deterministic by allowing the
scheduler to delay its next executed thread at most $k$ times. 
The bound $k$ is chosen independently of the number of threads. This
technique has been used for the analysis and testing of concurrent
programs~\cite{DBLP:conf/lopstr/MusuvathiQ06}.

SAT/SMT-based BMC has also been applied to the verification
of multi-threaded programs. 
In~\cite{DBLP:conf/cav/RabinovitzG05} a SAT-based BMC that bounds the
number of context switches has been described. In this work, for each
thread, a set of constraints describing the thread is generated using
BMC techniques for sequential programs~\cite{DBLP:conf/tacas/ClarkeKL04}. 
Constraints for concurrency describing both the number of context
switches and the reading or writing of global variables are then added
to the previous sets of constraints.
The work in~\cite{DBLP:conf/spin/GanaiG08} is also concerned with
efficient modeling of multi-threaded programs using SMT-based BMC.
Unlike~\cite{DBLP:conf/cav/RabinovitzG05}, in this work the
constraints for concurrency are added lazily during the BMC unrolling.

\subsection{Verification via Sequentialization}
\label{subsec:Sequentialization}

Yet another approach used for verifying multi- threaded programs is by
reducing bounded concurrent analysis to sequential analysis. In this
approach the multi-threaded program is translated into a sequential
program such that the latter over-approximates the bounded
reachability of the former. The resulting sequential program can then
be analyzed using any existing model checker for sequential programs.

This approach has been pioneered by the work
in~\cite{DBLP:conf/pldi/QadeerW04}. In this work a multi-threaded
program is converted to a sequential one that simulates all the
interleavings generated by multiple stacks of the multi-threaded
program using its single stack. The simulation itself is bounded by
the size of a multiset that holds existing runnable threads at any
time during the execution of a thread.

Lal and Reps~\cite{DBLP:journals/fmsd/LalR09} propose a translation
from multi-threaded programs to sequential programs that reduces the
context-bounded reachability of the former to the reachability of the
latter for any context bound.
Given a bound $k$, the translation constructs a sequential program
that tracks, at any point, only the local state of one thread, the
global state, and $k$ copies of the global state. 
In the translation each thread is processed separately from the
others, and updates of global states caused by context switches in the
processed thread are modeled by guessing future states using prophecy
variables and constraining these variables at an appropriate control
point in the execution. Due to the prophecy variables, the resulting
sequential program explores more reachable states than that of the
original multi-threaded program.
A similar translation has been proposed in~\cite{DBLP:conf/cav/TorreMP09}.
But this translation requires the sequential program to call the
individual thread multiple times from scratch to recompute the local
states at context switches.

As shown in Section~\ref{sec:Application}, and also
in~\cite{DBLP:conf/fmcad/CimattiMNR10}, the verification of
multi-threaded programs via sequentialization and abstraction-based
software model checking techniques turns out to suffer from several
inefficiencies. 
First, the encoding of the scheduler makes the sequential program more
complex and harder to verify.
Second, details of the scheduler are often needed to verify the
properties, and thus abstraction-based technique requires many
abstraction-refinement iterations to re-introduce the abstracted
details.

\subsection{Partial-Order Reduction}
\label{subsec:PORRelated}

\POR is an effective technique for reducing the search space by
avoiding visiting redundant executions. It has been mostly adopted in
explicit-state model checkers, like
\spin~\cite{DBLP:journals/ac/Holzmann05,HolzmannDoronFDT95,DBLP:journals/fmsd/Peled96},
\verisoft~\cite{DBLP:journals/fmsd/Godefroid05}, and
\zing~\cite{DBLP:conf/cav/AndrewsQRRX04}.
Despite their inability to handle infinite-domain inputs, the
maturity of these model checkers, in particular the support for \POR,
has attracted research on encodings of multi-threaded programs into
the language that the model checkers accept. 
In~\cite{DBLP:conf/spin/CampanaCNR11} we verify \systemc models by
encoding them in \promela, the language accepted by the
\spin model checker. The work shows that the resulting encodings lose
the intrinsic structures of the multi-threaded programs that are
important to enable optimizations like \POR.

There have been several attempts at applying \POR to symbolic model
checking techniques~\cite{DBLP:journals/fmsd/AlurBHQR01,DBLP:conf/cav/KahlonGS06,DBLP:conf/tacas/WangYKG08}.
In these applications \POR 
is achieved by statically adding constraints describing the reduction
technique into the encoding of the program.
The work in~\cite{DBLP:journals/fmsd/AlurBHQR01} apply \POR technique 
to symbolic BDD-based invariant checking.
While the work in~\cite{DBLP:conf/tacas/WangYKG08} describes an approach
that can be considered as a symbolic sleep-set based technique.
They introduce the notion of guarded independence relation, where a
pair of transitions are independent of each other if certain
conditions specified in the pair's guards are satisfied.
The \POR techniques applied into \ESST can be extended to use 
guarded independence relation by exploiting the thread and global regions.
Finally, the work in~\cite{DBLP:conf/cav/KahlonGS06} uses patterns of lock
acquisition to refine the notion of independence transition, which subsequently 
yields better reductions.

%% file: conclusions.tex
\section{Conclusions and Future Work}
\label{sec:ConclusionFutureWork}

In this paper we have presented a new technique, called \ESST, for the
verification of shared-variable multi-threaded programs with
cooperative scheduling.
The \ESST algorithm uses explicit-state model checking techniques to
handle the scheduler, while analyzes the threads using symbolic
techniques based on lazy predicate abstraction. Such a combination
allows the \ESST algorithm to have a precise model of the scheduler,
to handle it efficiently, and also to benefit from the effectiveness
of explicit-state techniques in systematic exploration of thread
interleavings. At the same time, the use of symbolic techniques allows
the \ESST algorithm to deal with threads that potentially have
infinite state space. 
\ESST is futher enhanced with \POR techniques, that prevents the
exploration of redundant thread interleavings.
The results of experiments carried out on a general class of
benchmarks for \systemc and \fairthreads cooperative threads clearly
shows that \ESST outperforms the verification approach based on
sequentialization, and that \POR can effectively improve the
performance.

As future work, we will proceed along different directions.
We will experiment with lazy abstraction with interpolants~\cite{DBLP:conf/cav/McMillan06},
to improve the performance of predicate abstraction
when there are too many predicates to keep track of.
We will also investigate the possibility of applying symmetry
reduction~\cite{DBLP:conf/cav/DonaldsonKKW11} to deal with cases where
there are multiple threads of the same type, and possibly with
parametrized configurations.

We will extend the \ESST algorithm to deal with primitive
function calls whose arguments cannot be inferred statically. This
requires a generalization of the scheduler exploration with a hybrid
(explicit-symbolic or semi-symbolic) approach, and the use of SMT
techniques to enumerate all possible next states of the scheduler.
Finally, we will look into the possibility of applying the \ESST
algorithm to the verification of general multi-threaded programs. This
work amounts to identifying important program locations in threads where
the control \emph{must} be returned to the scheduler.

%% file: appendix.tex
\section{Proofs of Lemmas and Theorems.}
\label{sec:appendixProofs}

\begin{LemStar}[\ref{lem:Overapprox}]
Let $\frnode$ and $\frnode'$ be \ARF nodes for a threaded program $P$
such that $\frnode'$ is a successor node of $\frnode$.  Let $\cg$ be a
configuration of $P$ such that $\cg \models \frnode$.  The following
properties hold:
\begin{enumerate}[\em(1)]
  \item If $\frnode'$ is obtained from $\frnode$ by 
  the rule~\ref{rule:ARFExp1} with the performed operation $op$, then,
  for any configuration $\cg'$ of $P$ such that $\cg \transsem{op} \cg'$,
  we have $\cg' \models \frnode'$.
  
  \item If $\frnode'$ is obtained from $\frnode$ by 
  the rule~\ref{rule:ARFExp2}, then, for any configuration $\cg'$ of $P$ 
  such that $\cg \transsem{\cdot} \cg'$ and the scheduler states 
  of $\frnode'$ and $\cg'$ coincide, we have $\cg' \models \frnode'$.
\end{enumerate}
\end{LemStar}
\ProofLemOverapprox

\begin{ThmStar}[\ref{thm:CorrectESST}]
Let $P$ be a threaded program. For every terminating execution of
$\ESST(P)$, we have the following properties:
\begin{enumerate}[\em(1)]
  \item If $\ESST(P)$ returns a feasible counter-example path $\fpath$, 
        then we have $\cg \simtrans{\fpath} \cg'$ for an initial
        configuration $\cg$ and an error configuration $\cg'$ of $P$.
  \item If $\ESST(P)$ returns a safe \ARF $\forest$, then for every 
        configuration $\cg \in \ReachC{P}$, there is an \ARF node
        $\frnode \in \NodesF{\forest}$ such that $\cg \models
        \frnode$.
\end{enumerate}
\end{ThmStar}
\ProofThmCorrectESST

\begin{ThmStar}[\ref{thm:PORCorrect}]
A transition system $M=(S,S_0,T)$ is safe w.r.t. a set $\Terr
\subseteq T$ of error transitions iff $\ReachRed{S_0}{T}$ that
satisfies the cycle condition does not contain any error state from
$\Err{M}{\Terr}$. 
\end{ThmStar} 
\ProofThmPORCorrect

\begin{LemStar}[\ref{lem:CommutePreserve}]
Let $\transa$ and $\transb$ be transitions that are independent of
each other such that for concreate states $s_1,s_2,s_3$ and abstract
state $\frnode$ we have $s_1 \models \frnode$, and both
$\transaSS{s_1}{s_2}$ and $\transbSS{s_2}{s_3}$ hold.
Let $\frnode'$ be the abstract successor state of $\frnode$ by
applying the abstract strongest post-operator to $\frnode$ and
$\transb$, and $\frnode''$ be the abstract successor state of
$\frnode'$ by applying the abstract strongest post-operator to
$\frnode'$ and $\transa$.
Then, there are concrete states $s_4$ and $s_5$ such that:
$\transbSS{s_1}{s_4}$ holds, 
$s_4 \models \frnode'$, 
$\transbSS{s_4}{s_5}$ holds, 
$s_5 \models \frnode''$, and 
$s_3 = s_5$.
\end{LemStar}
\ProofLemCommutePreserve

\begin{ThmStar}[\ref{thm:CorrectPORESST}]
Let $P$ be a threaded sequential program. For every
terminating executions of $\ESST(P)$ and $\ESST_{\POR}(P)$, we have
that $\ESST(P)$ reports safe iff so does $\ESST_{\POR}(P)$. 
\end{ThmStar} 
\ProofThmPORESSTCorrect